\documentclass[aps,twocolumn,showpacs,superscriptaddress,amssymb]{revtex4-1}
\date{\today}
\usepackage{graphicx,threeparttable,amssymb}
\usepackage{dcolumn}
\usepackage{bm}
\usepackage{color}

\begin{document}

\title{Systematic shell-model study on spectroscopic properties in the south region of $^{208}$Pb}

\author{Cenxi Yuan}
\email{yuancx@mail.sysu.edu.cn} \affiliation{Sino-French Institute
of Nuclear Engineering and Technology, Sun Yat-Sen University,
Zhuhai, 519082, Guangdong, China}

\author{Menglan Liu}
\affiliation{Sino-French Institute of Nuclear Engineering and Technology, Sun Yat-Sen University, Zhuhai, 519082, Guangdong, China}

\author{Noritaka Shimizu}
\affiliation{Center for Nuclear Study, University of Tokyo, Hongo, Bunkyo-ku, Tokyo 113-0033, Japan}

\author{Zs. Podoly\'{a}k}
\affiliation{Department of Physics, University of Surrey, Guildford, Surrey GU2 7XH, United Kingdom}

\author{Toshio Suzuki}
\affiliation{Department of Physics, College of Humanities and
Sciences, Nihon University, Sakurajosui 3, Setagaya-ku, Tokyo
156-8550, Japan}
\affiliation{National Astronomical Observatory of
Japan, Mitaka, Tokyo 181-8588, Japan }

\author{Takaharu Otsuka}
\affiliation{Department of Physics, University of Tokyo, Hongo, Bunkyo-ku, Tokyo 113-0033, Japan}
\affiliation{RIKEN Nishina Center, 2-1 Hirosawa, Wako, Saitama 351-0198, Japan}
\affiliation{National Superconducting Cyclotron Laboratory, Michigan
State University, East Lansing, Michigan, 48824, USA}
\affiliation{Instituut voor Kern-en Stralingsfysica, Katholieke Universiteit Leuven, B-3001 Leuven, Belgium}

\author{Zhong Liu}
\affiliation{Institute of Modern Physics, Chinese Academy of Sciences, Lanzhou 730000, China}
\affiliation{School of Nuclear Science and Technology, University of Chinese Academy of Sciences, Beijing 100049, China}

\begin{abstract}
  \begin{description}
  \item[Background]
    The properties of nuclei located in the south region of $^{208}$Pb are
    important for understanding the $r$-process nucleosynthesis.
    While some isomeric states and their spectroscopic properties have been investigated experimentally
    in neutron-rich Pb, Tl, and Hg isotopes recently,
    a large portion of the area still remains unreachable.

  \item[Purpose]
    We aim to study the properties of nuclei in the south region of $^{208}$Pb systematically,
    including the binding and excitation energies and electromagnetic properties,
    in order to predict unknown properties of these nuclei, such as isomerism,
    utilizing a theoretical model which describes the experimentally known properties precisely.
    We also address whether the $N=126$ shell closure is robust or not
    when the proton number decreases from $^{208}$Pb.

  \item[Methods]
    We performed large-scale shell-model calculations with a new Hamiltonian suggested in the present work.
    The model space is taken as the five proton orbits within $50<Z\leqslant82$ and the thirteen neutron orbits within $82<N\leqslant184$.
    And one-particle one-hole excitation is allowed across the $N=126$ gap.
    The Hamiltonian is constructed by combining the existing Hamiltonians, KHHE (with adjustment of its proton-proton part) and KHPE, and the monopole
    based universal interaction.  

  \item[Results] The shell-model results well reproduce the experimentally observed binding energies and spectroscopic properties, such as isomerism, core excitation, and electromagnetic properties.
    Some possible isomeric states in neutron-rich Pb, Tl, and Hg isotopes are predicted
    with transition energies and half-lives.
    The $N = 126$ shell gap is predicted to be robust from $Z = 82$ down to $68$ with minor reduction.
    We also examine the effective charges and the quenching of the $g$ factors suitable for this region
    by systematic comparisons between observed and calculated electromagnetic properties.

  \item[Conclusions]
    A new Hamiltonian is constructed for nuclei in the south region of $^{208}$Pb,
    mainly including Pb, Tl, Hg, Au, Pt, Ir, Os, Re, and W isotopes around $N=126$,
    and provides them reasonable descriptions on nuclear properties including
    binding energies, excitation energies and electromagnetic properties
    through comprehensive and systematic studies. The present Hamiltonian and discussions provide fruitful information for future measurements and theoretical investigations for nuclei in this region, especially those around the $N=126$ shell, including the recommended effective charges and $g$ factors,the predicted binding energies, isomeric states, and core-excited states.

  \end{description}
\end{abstract}


\maketitle

\section{\label{sec:level1}Introduction}
The rapid neutron-capture process (\emph{r} process) is believed to be a key origin of the heavy nuclides found in nature~\cite{Mumpower2016}.
The properties of nuclei, such as masses, half-lives, and neutron capture cross sections, are essential for understanding the \emph{r}-process nucleosynthesis~\cite{Kajino2017}. For example, the $A = 195$ abundance peak in the \emph{r} process is associated with the properties of nuclei around $N = 126$ ~\cite{Mumpower2016}. However, the measurements of many important nuclear data are beyond the present experimental access. Theoretical predictions of the unknown data of $N = 126$ isotones below $^{208}$Pb and nearby nuclei are of crucial significance.

Some isomers have been discovered in the region around $^{208}$Pb~\cite{Steer2011}, but many more are expected from shell-model calculations. On the neutron-rich side of the stability line in this region, the $8^{+}_{1}$ seniority isomers have been observed up to $^{216}$Pb and $^{210}$Hg in their respective isotopic chains~\cite{Gottardo2012,Gottardo2013}. The KISS (KEK Isotope Separation System) project aims to measure the properties of the ground and isomeric states in the south region of $^{208}$Pb \cite{Miyatake2018,Watanabe2015}, such as the magnetic moments of the ground and isomeric states of $^{199}$Pt \cite{Hirayama2017}. Reliable descriptions of the known levels and electromagnetic transition rates are necessary for a model to make new predictions in this region. The $\beta$-decay properties of long-lived isomers in this region could also be relevant to network calculations in the \emph{r}-process nucleosynthesis.

Also, the stability of $N=126$ shell closure in the neutron-rich side attracts much interest. In nuclei with extreme proton-neutron ratio, traditional proton (neutron) magic numbers may disappear and new ones may emerge, manifesting dynamic shell evolution with isospin. Recent investigations revealed a significant $sd$ component in both the ground and excited states in $^{12}$Be~\cite{Chen2018-1,Chen2018-2}, indicating the disappearance of the $N=8$ shell closure in very neutron-rich Be isotopes. The $N=20$ shell closure vanishes below $^{34}$Si, the northern boundary of the ``island of inversion''~\cite{Han2017}. In medium mass nuclei, the $Z=40$ sub-shell in neutron-rich Ag isotopes with neutron numbers approaching $N=82$ are found very fragile \cite{Chen2019}.

So far, experimental information is scarce on the size of the $N=126$ shell gap in the most neutron-rich isotopes with $Z\leq82$. An observation of the excited states in $^{204}$Pt supported an unquenched $N = 126$ gap~\cite{Steer2008}. Some neutron core-excited states are found at excitation energies larger than $2$ MeV in $^{208,209}$Pb, $^{207}$Tl, and $^{206}$Hg~\cite{nndc,Wilson2015}. At present, it is still challenging to observe neutron core-excited states in $N=126$ isotones below $^{206}$Hg. It is necessary to construct a theoretical model, which describes the experimentally available data of $^{207}$Tl and $^{206}$Hg precisely, for a reliable prediction of the evolution of the $N=126$ shell gap with decreasing atomic number $Z$.

The nuclear shell model is one of the best models to describe the above structure properties in a unified way, including binding energies, levels, shell gaps, electromagnetic transition rates, isomers, and $\beta$ decays. The present work aims to provide a systematic shell-model investigation based on a newly constructed Hamiltonian for the description of spectroscopic properties in the region around $^{208}$Pb. The properties mentioned above are discussed except for $\beta$ decays. The first forbidden transitions of $\beta$ decays are of crucial importance for describing the decay half-lives and delayed neutron emissions from heavy nuclei~\cite{Suzuki2012,Nishimura2016}.

The present paper is organized as follows: in Sec. \ref{sec:level2}, a brief description of the construction of the Hamiltonian is given. Section \ref{sec:level3} presents the results of binding energies. The low-lying levels and core-excited properties, including the evolution of the $N=126$ shell gap, are discussed in Secs. \ref{sec:level4} and \ref{sec:level5}, respectively. Section \ref{sec:level6} and \ref{sec:level7} present discussions on electric quadrupole and magnetic dipole properties, respectively. A summary is given in Sec. \ref{sec:level8}.

\section{\label{sec:level2}Shell-Model Hamiltonian}

\begin{table}
\caption{\label{H} Construction of the two-body part of the present Hamiltonian.}
\begin{ruledtabular}
\begin{tabular}{ccc}
        & type of interaction         & source of interaction \\ \hline
PO5-PO5 & proton-proton   & KHHE (modified) \\
NO6-NO6 & neutron-neutron & KHHE \\
NO7-NO7 & neutron-neutron & KHPE \\
PO5-NO6 & proton-neutron  & KHHE \\
PO5-NO7 & proton-neutron  & V$_{\text{MU}}$+LS \\
NO6-NO7 & neutron-neutron & V$_{\text{MU}}$+LS \\
\end{tabular}
\end{ruledtabular}
\end{table}

The present Hamiltonian is constructed in the model space with five proton orbits and thirteen neutron orbits. The five proton orbits, $0g_{7/2}$, $1d_{5/2}$, $1d_{3/2}$, $2s_{1/2}$, and $0h_{11/2}$, are named PO5 (5 proton orbits) in the following discussions. The six neutron orbits below the $N = 126$ shell gap, $0h_{9/2}$, $1f_{7/2}$, $1f_{5/2}$, $2p_{3/2}$, $2p_{1/2}$, and $0i_{13/2}$, are named NO6 (6 neutron orbits) in the following discussions.
The seven neutron orbits above the $N = 126$ shell gap, $0i_{11/2}$, $1g_{9/2}$, $1g_{7/2}$, $2d_{5/2}$, $2d_{3/2}$, $3s_{1/2}$, and $0j_{15/2}$, are named NO7. The present model space does not include the proton orbits above the $Z=82$ shell gap, because many core-excited states of $^{208}$Pb, $^{207}$Tl, and $^{206}$Hg are dominated by one-neutron core-excited configuration, as discussed in Sec. \ref{sec:level5}.

The single-particle energies of the present Hamiltonian are fixed to reproduce the single-particle levels of $^{207,209}$Pb and $^{207}$Tl, the proton separation energy of $^{208}$Pb, and the neutron separation energies of $^{208,209}$Pb~\cite{nndc,AME2020}.

The Hamiltonian KHHE~\cite{Warburton1991-1} was constructed in the model space consisting of PO5 and NO6. KHHE is included in the present Hamiltonian for the proton-proton interaction inside PO5, the neutron-neutron interaction inside NO6, and the proton-neutron interaction between PO5 and NO6. KHHE is based on the Kuo-Herling hole (KHH) interaction \cite{Kuo1971,Herling1972}. As suggested in Ref.~\cite{Rydstrom1990}, proton-proton and proton-neutron interactions of KHH are modified to give more precise descriptions of the levels in nuclei around $^{208}$Pb. The construction of KHHE included such modifications and further considered modifications on neutron-neutron interaction~\cite{Warburton1991-1}.

The Hamiltonian KHPE~\cite{Warburton1991-2}, which is based on Kuo-Herling particle (KHP) interaction \cite{Kuo1971,Herling1972}, is constructed in the model space including six proton orbits beyond $Z = 82$ and NO7. Recently, KHPE is used to analyze the spin and parity of newly observed nuclide $^{223}$Np~\cite{sun2017PLB} and isomeric state of $^{218}$Pa~\cite{zhang2020PLB}. The neutron-neutron part of KHPE is used for the neutron-neutron interaction inside NO7 in the present Hamiltonian.

In the present Hamiltonian, the monopole based universal interaction V$_{\text{MU}}$ \cite{otsuka2010} and the spin-orbit force from M3Y~\cite{m3y1977}(V$_{\text{MU}}$+LS) are used for the proton-neutron interaction between PO5 and NO7 and the neutron-neutron interaction between NO6 and NO7. The V$_{\text{MU}}$+LS interaction was examined to be suitable for shell-model calculations in different regions, such as the $psd$ region~\cite{yuan2012}, $sdpf$~\cite{utsuno2012}, $pfsdg$ region~\cite{Togashi2015}, the southeastern region of $^{132}$Sn~\cite{yuan2016}, and the northwestern region of  $^{208}$Pb~\cite{zhang2021-U214, yang2022-Th207}.

Table \ref{H} summarizes the construction of the Hamiltonian. Because the proton-proton interaction inside PO5 is from KHHE which includes the Coulomb interaction, the calculated binding energies do not need Coulomb correction. In addition, $0.1$MeV is added to all two-body matrix elements (TBME) of proton-proton interaction except $\langle2s_{1/2}2s_{1/2}\mid$V$\mid2s_{1/2}2s_{1/2}\rangle$. Such modification provides more precise descriptions of the binding energies of Au, Pt, and Ir isotopes, which will be discussed in Sec. \ref{sec:level3}. The modification rarely affects the spectroscopic properties of Pb, Tl, and Hg isotopes, because the monopole effect is canceled in these isotopes when the single-particle energies of these proton orbits are fixed to those of $^{207}$Tl.

Systematic shell-model calculations are challenging in the full medium and heavy mass region due to the huge computation dimension of most of nuclei. As the computation consumption increases exponentially with the valence particles (holes), this work mainly focuses on the Pb, Tl, and Hg isotopes and the $A\geq198$ nuclei, around the $N = 126$ shell. As for the nuclei around $A = 190$ whose properties have been experimentally measured, are anticipated to be investigated through approximations, such as further truncation on the model space.

The consideration of core-excitation also extends the computation dimension. Furthermore, considering the mixing between the normal states and the core-excited states would aggravate the case and introduces many complicated correlations. For example, if one neutron crossing $N = 126$ shell is considered, the diagonal TBME $\langle$(PO5)(NO7)$\mid$V$\mid$(PO5)(NO7)$\rangle$ and  $\langle$(NO6)(NO7)$\mid$V$\mid$(NO6)(NO7)$\rangle$ should be included. But if the mixing between the normal states and the core-excited states are considered, more off-diagonal TBME $\langle$(PO5)(NO6)$\mid$V$\mid$(PO5)(NO7)$\rangle$ and  $\langle$(NO6)(NO6)$\mid$V$\mid$(NO6)(NO7)$\rangle$ should be considered. Whereas, the properties and strength of the off-diagonal cross-shell interaction have been rarely studied. It is shown that a weaker off-diagonal cross-shell interaction can reproduce the low lying properties of $^{14}$C in a 4$\hbar\omega$ calculation~\cite{yuan2017}. But it is more complicated in heavy nuclei because there are orbits with three major shell differences, such as $0g_{7/2}$ and $0j_{15/2}$ orbits. As a result, only one-neutron excitation across the $N = 126$ shell gap is considered for core-excited states in the present work.

\begin{table*}
\caption{\label{BE1-1} The comparison between the calculated and observed binding energies (unit in MeV). Experimental data are taken from AME2020~\cite{AME2020}. Root mean square (RMS) deviation between calculations and observations is for all $56$ nuclei.}
\begin{ruledtabular}
\begin{tabular}{cccc cccc cccc}
nuclide&BE$_{SM}$&BE$_{Expt.}$&$\Delta$BE&nuclide&BE$_{SM}$&BE$_{Expt.}$&$\Delta$BE&nuclide&BE$_{SM}$&BE$_{Expt.}$&$\Delta$BE\\ \hline
$^{215}$Pb & 1666.803 & 1666.839 & 0.036  &$^{211}$Tl & 1645.832 & 1645.756 & -0.076 & $^{203}$Hg & 1601.067 & 1601.159 &  0.092   \\
$^{214}$Pb & 1663.220 & 1663.293 & 0.073  &$^{210}$Tl & 1640.891 & 1640.854 & -0.037 & $^{202}$Hg & 1595.141 & 1595.164 &  0.023  \\
$^{213}$Pb & 1658.194 & 1658.241 & 0.047  &$^{209}$Tl & 1637.209 & 1637.180 & -0.029 & $^{201}$Hg & 1587.349 & 1587.410 &  0.061   \\
$^{212}$Pb & 1654.466 & 1654.516 & 0.050  &$^{208}$Tl & 1632.220 & 1632.214 & -0.006 & $^{200}$Hg & 1581.194 & 1581.179 &  -0.015  \\
$^{211}$Pb & 1649.367 & 1649.389 & 0.022  &$^{207}$Tl & 1628.427 & 1628.427 & 0.000  & $^{199}$Hg & 1573.149 & 1573.151 &  0.002  \\
$^{210}$Pb & 1645.522 & 1645.553 & 0.031  &$^{206}$Tl & 1621.582 & 1621.575 & -0.007 & $^{198}$Hg & 1566.690 & 1566.487 &  -0.203  \\
$^{209}$Pb & 1640.368 & 1640.368 & 0.000  &$^{205}$Tl & 1615.071 & 1615.071 & 0.000  & $^{203}$Au & 1599.855 & 1599.816 &  -0.039 \\
$^{208}$Pb & 1636.430 & 1636.430 & 0.000  &$^{204}$Tl & 1607.518 & 1607.525 & 0.007  & $^{202}$Au & 1592.925 & 1592.954 &  0.029 \\
$^{207}$Pb & 1629.062 & 1629.062 & 0.000  &$^{203}$Tl & 1600.930 & 1600.869 & -0.061 & $^{201}$Au & 1587.087 & 1586.930 &  -0.157  \\
$^{206}$Pb & 1622.323 & 1622.325 & 0.002  &$^{202}$Tl & 1593.021 & 1593.017 & -0.004 & $^{200}$Au & 1579.783 & 1579.698 &  -0.085 \\
$^{205}$Pb & 1614.291 & 1614.238 & -0.053 &$^{201}$Tl & 1586.283 & 1586.145 & -0.138 & $^{199}$Au & 1573.279 & 1573.481 &  -0.248 \\
$^{204}$Pb & 1607.564 & 1607.506 & -0.058 &$^{200}$Tl & 1578.049 & 1577.941 & -0.108 & $^{198}$Au & 1566.108 & 1565.896 &  -0.212 \\
$^{203}$Pb & 1599.222 & 1599.112 & -0.110 &$^{199}$Tl & 1571.142 & 1570.882 & -0.260 & $^{202}$Pt & 1591.962 & 1592.075 &  0.113  \\
$^{202}$Pb & 1592.319 & 1592.195 & -0.124 &$^{198}$Tl & 1562.569 & 1562.145 & -0.290 & $^{201}$Pt & 1585.121 & 1585.052 &  -0.069  \\
$^{201}$Pb & 1583.648 & 1583.454 & -0.194 &$^{208}$Hg & 1629.480 & 1629.512 &  0.032 & $^{200}$Pt & 1579.937 & 1579.840 &  -0.097 \\
$^{200}$Pb & 1576.602 & 1576.362 & -0.240 &$^{207}$Hg & 1624.618 & 1624.662 &  0.044 & $^{199}$Pt & 1572.711 & 1572.558 &  -0.153 \\
$^{199}$Pb & 1567.569 & 1567.272 & -0.297 &$^{206}$Hg & 1620.993 & 1621.049 &  0.056 & $^{198}$Pt & 1567.271 & 1567.002 &  -0.269 \\
$^{198}$Pb & 1560.423 & 1560.036 & -0.387 &$^{205}$Hg & 1614.239 & 1614.320 &  0.081 & $^{199}$Ir & 1570.561 & 1570.351 &  -0.210 \\
$^{213}$Tl & 1654.249 & 1654.037 & -0.212 &$^{204}$Hg & 1608.548 & 1608.651 &  0.103 & RMS        &          &          &  0.134 \\

\end{tabular}
\end{ruledtabular}
\end{table*}

\begin{table*}
\caption{\label{BE1-2} The predicted binding energies from the present Hamiltonian and AME2020~\cite{AME2020} (unit in MeV).}
\begin{ruledtabular}
\begin{tabular}{cccccccccccc}
nuclide&BE$_{SM}$&BE$_{AME}$&$\Delta$BE&nuclide&BE$_{SM}$&BE$_{AME}$&$\Delta$BE&nuclide&BE$_{SM}$&BE$_{AME}$&$\Delta$BE \\ \hline
$^{217}$Pb & 1675.141 & 1675.023 &  -0.118  &  $^{206}$Pt & 1611.098 & 1610.920 &  -0.178  &   $^{200}$Os & 1573.440 & 1573.400 &  -0.040      \\
$^{216}$Pb & 1671.749 & 1671.840 &  0.091   &  $^{205}$Pt & 1606.562 & 1606.380 &  -0.182  &   $^{199}$Os & 1567.141 & 1566.926 &  -0.215   \\
$^{212}$Tl & 1649.379 & 1649.360 &  -0.019  &  $^{204}$Pt & 1603.291 & 1603.236 &  -0.055  &   $^{202}$Re & 1575.202 &          &           \\
$^{211}$Hg & 1641.145 & 1640.947 &  -0.198  &  $^{203}$Pt & 1597.105 & 1597.001 &  -0.104  &   $^{201}$Re & 1572.396 &          &           \\
$^{210}$Hg & 1637.801 & 1637.790 &  -0.011  &  $^{204}$Ir & 1596.142 & 1595.892 &  -0.250  &   $^{200}$Re & 1567.081 &          &           \\
$^{209}$Hg & 1632.980 & 1632.917 &  -0.063  &  $^{203}$Ir & 1593.038 & 1592.535 &  -0.503  &   $^{199}$Re & 1562.566 & 1562.150 &  -0.416   \\
$^{209}$Au & 1627.867 & 1627.274 &  -0.593  &  $^{202}$Ir & 1587.182 & 1586.710 &  -0.472  &   $^{198}$Re & 1556.692 & 1556.478 &  -0.214         \\
$^{208}$Au & 1623.296 & 1623.024 &  -0.272  &  $^{201}$Ir & 1582.132 & 1581.870 &  -0.262  &   $^{201}$W  & 1564.982 &         &           \\
$^{207}$Au & 1619.925 & 1619.568 &  -0.357  &  $^{200}$Ir & 1575.699 & 1575.600 &  -0.099  &   $^{200}$W  & 1562.421 &          &           \\
$^{206}$Au & 1615.346 & 1615.040 &  -0.306  &  $^{198}$Ir & 1563.834 & 1563.606 &  -0.228  &   $^{199}$W  & 1557.215 &          &           \\
$^{205}$Au & 1611.820 & 1611.300 &  -0.520  &  $^{203}$Os & 1586.666 & 1586.242 &  -0.424  &   $^{198}$W  & 1553.076 &          &           \\
$^{204}$Au & 1605.441 & 1605.072 &  -0.369  &  $^{202}$Os & 1583.752 & 1583.478 &  -0.274  &   $^{197}$W  & 1547.252 & 1547.041 & -0.211    \\
$^{207}$Pt & 1614.225 & 1613.979 &  -0.246  &  $^{201}$Os & 1578.066 & 1577.649 &  -0.417  &              &          &          &    \\
\end{tabular}
\end{ruledtabular}
\end{table*}

\section{\label{sec:level3}Binding Energies}

Nuclear mass or binding energy is one of the most fundamental properties of an atomic nucleus. Many global mass models are available, such as the liquid drop model (LD)~\cite{myers1966}, finite-range droplet model~\cite{moller1995}, the Lublin-Strasbourg Drop model~\cite{pomorski2003}, the Hartree-Fock-Bogoliubov model~\cite{Goriely2009,Goriely2013}, and the Weizs\"{a}cker-Skyrme mass model~\cite{liu2011,Wang2014}. The overall uncertainties of the global mass models are around $0.5$ MeV. The recent version of Weizs\"{a}cker-Skyrme mass model includes the surface diffuseness effect for the extremely neutron-rich nuclei, where the neutrons may extend very far beyond the core~\cite{Wang2014}. The investigation on the neutron-halo nucleus, $^{22}$C, shows that its predicted radius is strongly affected by the neutron-neutron interaction between the two valence neutrons and the configurations of the $^{20}$C core~\cite{Suzuki2016}. There are some local mass models, which predict the unknown mass of a nucleus with high precision from the masses of the nearby nuclei, such as the Garvey-Kelson relations \cite{garvey1966} with generalizations \cite{cheng2014} and some new suggestions \cite{fu2018}.

Although the nuclear shell model is more often used to study the spectroscopic properties rather than the bulk properties, it can provide a precise description of masses or binding energies if the Hamiltonian is adequately tuned. Because the shell-model calculations are performed with a core, the calculated binding energy of a specific nucleus is the sum of the observed binding energy of the core and the shell-model binding energy of this nucleus relative to the core.

\begin{figure}
\includegraphics[scale=0.30]{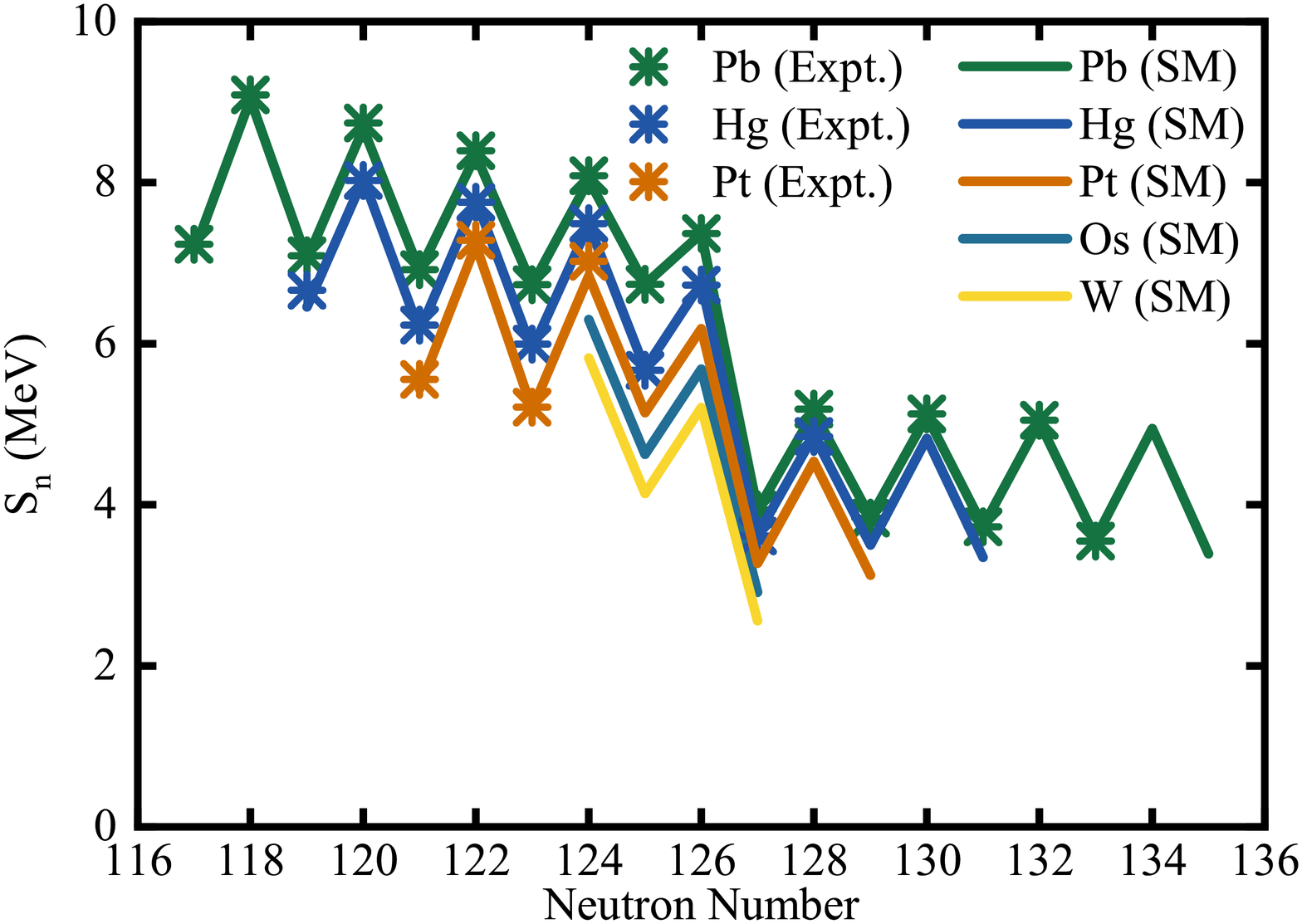}
\includegraphics[scale=0.30]{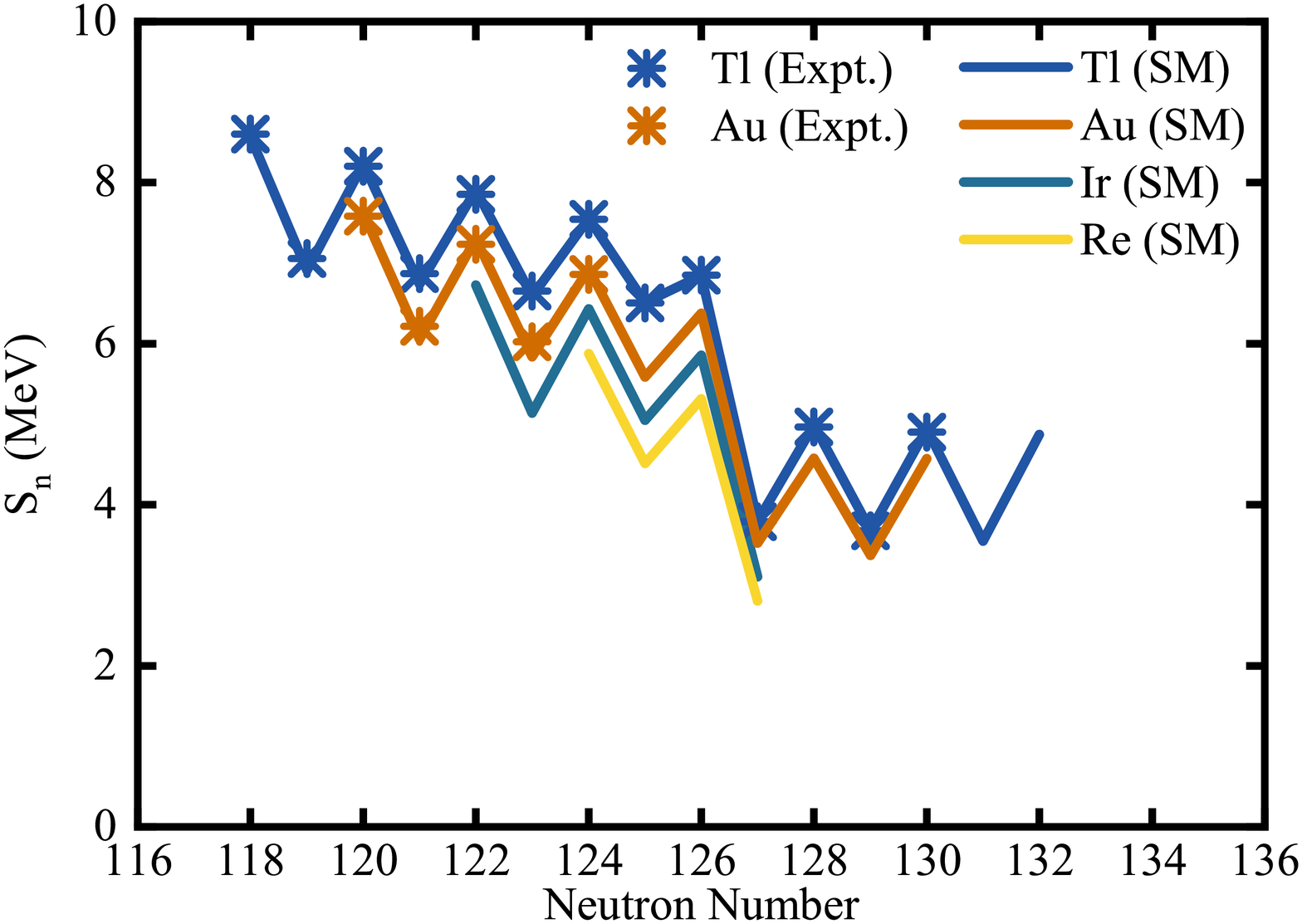}
\caption{\label{Sn} (Color online) The calculated and observed one neutron separation energies of Pb, Hg, Pt, Os, W (upper panel) and Tl, Au, Ir, Re (lower panel) isotopes. Experimental data are taken from AME2020~\cite{AME2020}. }
\end{figure}

As shown in Table \ref{BE1-1}, the binding energies calculated through the Hamiltonian described in the Sec. \ref{sec:level2}, labeled as BE$_{SM}$, are compared with the observed data. The $56$ experimental binding energies are well reproduced by the shell-model calculations with a root mean square (RMS) deviation of only $0.134$ MeV. Such a description is much more precise than the global mass models and comparable with the local mass models. In general, the nuclei close to $^{208}$Pb are more precisely reproduced. Based on the present Hamiltonian, some unmeasured binding energies are predicted in Table \ref{BE1-2} for neutron-rich Pb, Tl, Hg, Au, Pt, Ir, Os, Re, and W isotopes. For Ir, Os, Re, and W isotopes, only results of $N=123$, $124$, $125$, $126$, and $127$ isotones are presented because of the enormous computational cost for other isotones. In general, the present Hamiltonian provides similar predictions to those of AME2020, with an RMS deviation of $0.293$ MeV. Future mass measurements of these nuclei can further constrain the shell-model Hamiltonian.

If the proton-proton interaction is not modified, all binding energies of nuclei with $Z<80$, $^{198-203}$Au, $^{198-202}$Pt, and $^{199}$Ir, are overbound compared with the observed data, which are presented in appendix. The effect of the interaction modification is not apparent in the calculation of Pb, Tl, and Hg isotopes, because all proton orbits except $2s_{1/2}$ are almost fully occupied in these isotopes and their effect on binding energies are canceled when the single-particle energies are fitted to those of $^{207}$Tl. During the construction of the KHHE interaction, the experimental data of the Au, Pt, and Ir isotopes were not used for the fit. When the Au, Pt, and Ir isotopes are considered, the proton holes in other orbits contributes to the binding energies. Thus the proton $1d_{3/2}$, $1d_{5/2}$, $0g_{7/2}$, and $0h_{11/2}$ orbits are modified in the present work.

The observed and calculated one neutron separation energies ($S_{n}$) are presented in FIG.~\ref{Sn} for Pb, Tl, Hg, Au, Pt, Ir, Os, Re, and W isotopes. The present Hamiltonian almost exactly reproduces the observed $S_{n}$ values, including the odd-even staggering. The $S_{n}$ values show strong similarity among nine isotopes both above and below the $N=126$ shell gap, which indicates that the dominated neutron configurations are similar among these nuclei, as discussed in Sec. \ref{sec:level4}. Even in the neutron-rich Ir, Os, Re, and W isotopes around $N=126$, the predicted $S_{n}$ values are larger than $2.5$ MeV, which means that the neutron drip line is still very far from $N=126$ at $Z=74$.

Based on the excellent performance of the present Hamiltonian on the binding energies and separation energies, it is used to calculate the levels, cross-shell excitations, and electromagnetic properties of nuclei in the south region of $^{208}$Pb in the following discussions. 

\section{\label{sec:level4}Low-lying Levels}
The present Hamiltonian, with the modification on all $1d_{3/2}$, $1d_{5/2}$, $0g_{7/2}$, and $0h_{11/2}$ orbits, gives excitation energies in nice agreement with experimental values, including those of both low-lying and core-excited states, a prerequisite that the Hamiltonian can be used to predict isomers and shell closure around $N=126$. It is not necessary to discuss all calculated levels in the present work. Some states in neutron-rich Pb, Tl, and Hg isotopes are discussed in this section, which concentrates on possible isomeric states with small transition energies, such as the $8^{+}_{1}$ states in $^{210,212,214,216}$Pb, the $21/2^{+}_{1}$ and $27/2^{+}_{1}$ states in $^{211,213,215}$Pb, the $13/2^{+}_{1}$ and $17/2^{+}_{1}$ states in $^{209,211,213}$Tl. The half-lives of these possible isomeric states will be discussed in Sec. \ref{sec:level6}.

\subsection{$^{210,212,214,216}$Pb}

\begin{figure}
\includegraphics[scale=0.30]{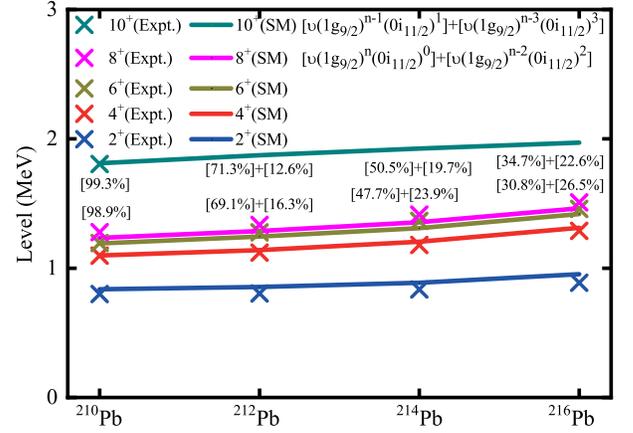}
\caption{\label{Pbeven} (Color online) The calculated and observed levels of $^{210,212,214,216}$Pb. Observed data are taken from NNDC~\cite{nndc}. The observed $8^{+}_{1}$ states in $^{214,216}$Pb are drawn 0.05 MeV above the corresponding $6^{+}_{1}$ states as estimations. Dominant configurations of the $8^{+}_{1}$ and $10^{+}_{1}$ states and their percentages are indicated.}
\end{figure}

As seen in FIG.~\ref{Pbeven}, all $8^{+}_{1}$ states in neutron-rich even-even Pb isotopes, $^{210,212,214,216}$Pb, are located just above the $6^{+}_{1}$ states, leading to the isomerism of these $8^{+}_{1}$ states with half-lives of hundreds of ns or several $\mu$s. The $8^{+}_{1}$ states in $^{214,216}$Pb are observed without exact decay energies to the $6^{+}_{1}$ states. As discussed in Ref.~\cite{Gottardo2012}, the energies of the $8^{+}_{1}$ to $6^{+}_{1}$ transitions in $^{214,216}$Pb are assumed to be between $0.02$ and $0.09$ MeV, which are drawn $0.05$ MeV above the $6^{+}_{1}$ states as estimations in FIG.~\ref{Pbeven}.

The dominant configuration of $0^{+}_{1}$, $2^{+}_{1}$, $4^{+}_{1}$, $6^{+}_{1}$, and $8^{+}_{1}$ in $^{210,212,214,216}$Pb is $\nu(1g_{9/2})^{n}$, where $n$ is the number of valence neutrons beyond the $N=126$ shell gap. The configuration $\nu(1g_{9/2})^{n-2}(0i_{11/2})^{2}$ plays a more important role when the neutron number increases. Because the energy differences $E(4^{+}_{1})-E(2^{+}_{1})$, $E(6^{+}_{1})-E(4^{+}_{1})$, and $E(8^{+}_{1})-E(6^{+}_{1})$ stay almost constant among $^{210,212,214,216}$Pb, these energy differences are mainly contributed by the coupling of two $\nu(1g_{9/2})$ neutrons.

The $10^{+}_{1}$ state was observed in $^{210}$Pb but not yet in $^{212,214,216}$Pb. It is interesting to discuss how $E(10^{+}_{1})$ varies when the neutron number increases. The calculations show $E(10^{+}_{1})$ in $^{212,214,216}$Pb are similar to that in $^{210}$Pb and much larger than $E(8^{+}_{1})$. The results indicate that the $10^{+}_{1}$ states are not from the pure $\nu(1g_{9/2})$ configuration but the $\nu(1g_{9/2})^{1}(0i_{11/2})^{1}$ configuration. The neutron occupancies of the $\nu(1g_{9/2})$ orbit are $3.4$, $4.8$, and $6.1$ in the $8^{+}_{1}$ states of $^{212,214,216}$Pb, respectively. Although many neutrons occupy this orbit, they do not tend to couple to an angular momentum $10$.

\subsection{$^{211,213,215}$Pb}

\begin{figure}
\includegraphics[scale=0.32]{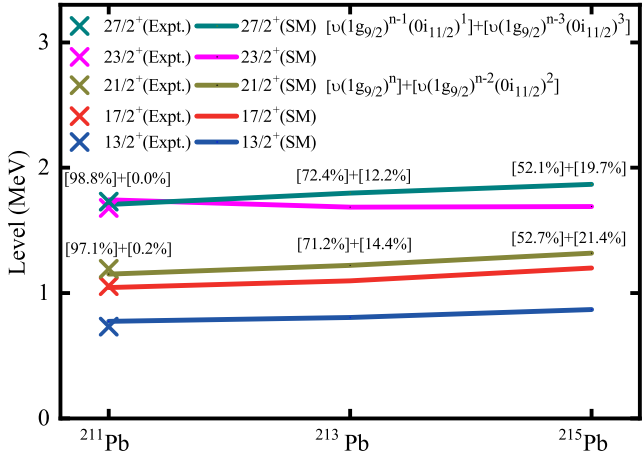}
\caption{\label{Pbodd} (Color online) The calculated and observed levels of $^{211,213,215}$Pb. Observed data are taken from NNDC~\cite{nndc}. The observed $27/2^{+}_{1}$ state in $^{211}$Pb is drawn 0.05 MeV above the corresponding $23/2^{+}_{1}$ state as an estimation. Dominant configurations of the $21/2^{+}_{1}$ and $27/2^{+}_{1}$ states and their percentages are indicated.}
\end{figure}

In $^{211}$Pb, the $21/2^{+}_{1}$ and $27/2^{+}_{1}$ states are located just above the $17/2^{+}_{1}$ and $23/2^{+}_{1}$ states, respectively, which are shown in FIG.~\ref{Pbodd}. Both of them are observed as isomeric states with half-lives of around one hundred ns. The shell-model results predict similar tendency in $^{213,215}$Pb.

The angular momenta of $17/2^{+}_{1}$ and $21/2^{+}_{1}$ are mainly from the coupling of three $\nu(1g_{9/2})$ neutrons, of which the maximum angular momentum is $21/2^{+}$. The angular momenta of $23/2^{+}_{1}$ and $27/2^{+}_{1}$ are mainly from the $\nu(1g_{9/2})^{2}(0i_{11/2})^{1}$ configuration, which can be considered as a $\nu(0i_{11/2})$ neutron coupled to the $8^{+}_{1}$ and $6^{+}_{1}$ states in the corresponding even Pb isotopes. Both $21/2^{+}_{1}$ and $27/2^{+}_{1}$ in $^{213,215}$Pb are possible isomeric states through the present calculations. But the energy gaps between the $27/2^{+}_{1}$ and $23/2^{+}_{1}$ states become larger from $^{211}$Pb to $^{215}$Pb, which leads to shorter half-lives of the $27/2^{+}_{1}$ states.

\subsection{$^{209,211,213}$Tl}

\begin{figure}
\includegraphics[scale=0.30]{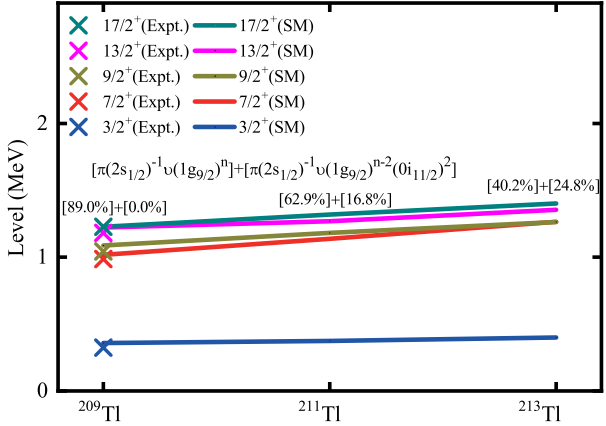}
\caption{\label{Tlodd} (Color online) The calculated and observed levels of $^{209,211,213}$Tl. Observed data are taken from NNDC~\cite{nndc} and Ref.~\cite{Amro2017}. Dominant configurations of the $17/2^{+}_{1}$ states and their percentages are indicated.}
\end{figure}

As shown in FIG.~\ref{Tlodd}, the $17/2^{+}_{1}$ state is located just above the $13/2^{+}_{1}$ state and is an isomeric state in $^{209}$Tl. The excitation energies are rather constant among $^{209,211,213}$Tl. The $17/2^{+}_{1}$ and $13/2^{+}_{1}$ states in $^{211,213}$Tl are possible isomeric states. All $9/2^{+}_{1}$, $13/2^{+}_{1}$, and $17/2^{+}_{1}$ states are dominated by the same configuration $\pi(2s_{1/2})^{-1}\nu(1g_{9/2})^{2}$, which can be considered as a $\pi(2s_{1/2})$ proton hole coupled to the $8^{+}_{1}$ and $6^{+}_{1}$ states in the corresponding Pb isotopes. The $9/2^{+}_{1}$ states are above the $7/2^{+}_{1}$ states in $^{209,211}$Tl, which decay through both M1 and E2 transitions. Although the $9/2^{+}_{1}$ state becomes lower than the $7/2^{+}_{1}$ state in $^{213}$Tl, it is not predicted to be an isomeric state because there is a $5/2^{+}_{1}$ state at $0.4$ MeV lower.

The present structure agrees well with the recent observed $0.58(8)$ $\mu$s isomer in $^{211}$Tl~\cite{Gottardo2019}, which suggests that the $17/2^{+}_{1}$ state is the isomer. Our prediction on the half-life of the $17/2^{+}_{1}$ state in $^{211}$Tl is $0.65$ $\mu$s, which will be shown in Sec.~\ref{sec:level6}.

Two isomers de-exciting with gamma transitions of $380$ and $698$ keV, respectively, are found in $^{213}$Tl~\cite{Gottardo2019}. Both the calculated positive-parity states in Ref.~\cite{Gottardo2019} and the present work do not seem to explain these two isomers. The isomer decaying via the $380$ keV transition may correspond to the $11/2^{-}_{1}$ state with a proton $0h_{11/2}$ hole configuration. The $11/2^{-}_{1}$ state is calculated to be just $10$ keV above the $9/2^{+}_{1}$ state. Because of the uncertainty of calculation, $11/2^{-}_{1}$ may be actually below the $9/2^{+}_{1}$ state and decay to the $5/2^{+}_{1}$ state with decay energy $422$ keV. The $698$ keV $\gamma$ line is rather weak and not discussed here.

\subsection{$^{208,210,212}$Tl}

\begin{figure}
\includegraphics[scale=0.30]{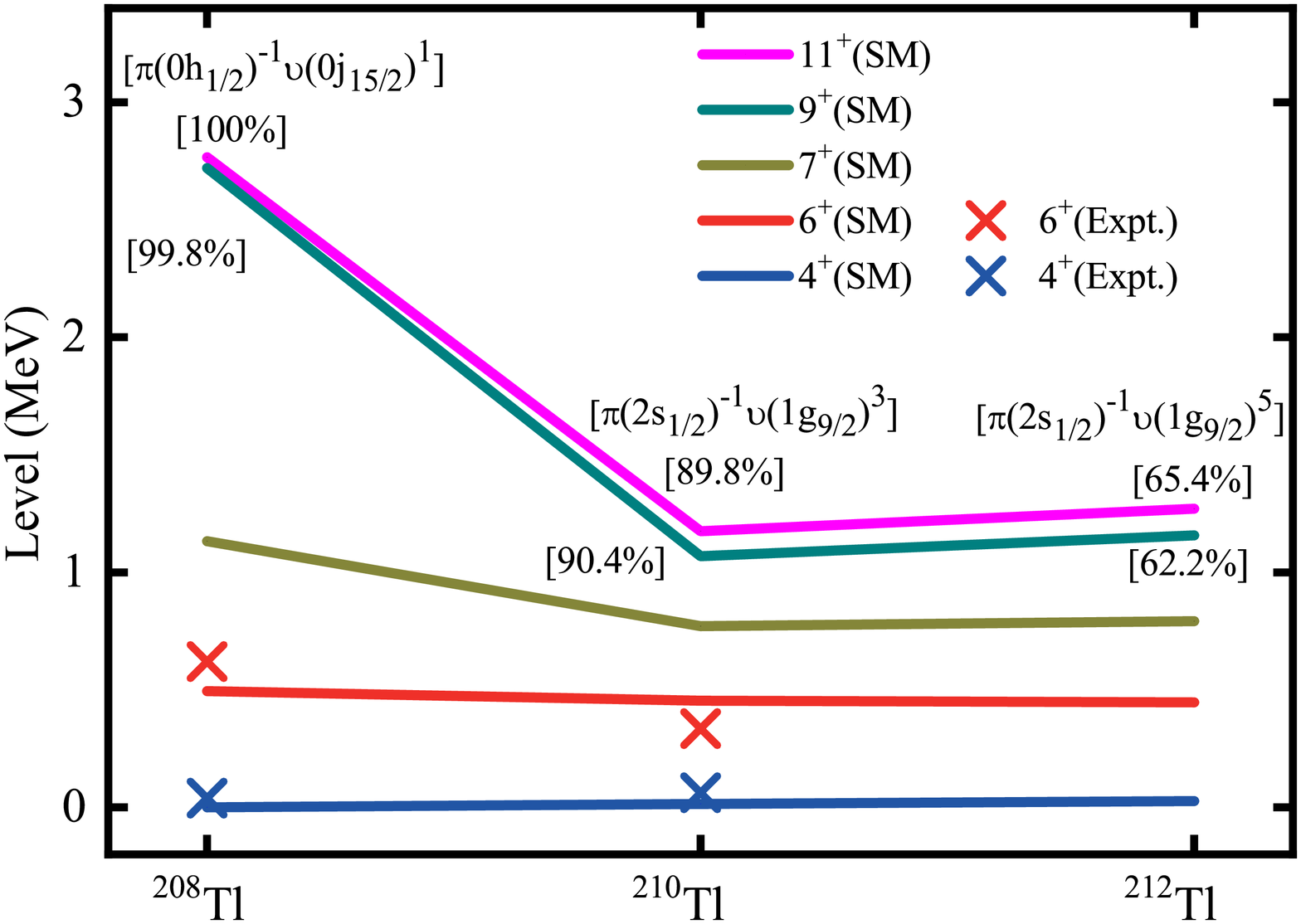}
\caption{\label{Tleven} (Color online) The calculated and observed levels of $^{208,210,212}$Tl. Observed data are taken from NNDC~\cite{nndc}. Dominant configurations of the $9^{+}_{1}$ and $11^{+}_{1}$ states and their percentages are indicated.}
\end{figure}

As shown in FIG.~\ref{Tleven}, the $11^{+}_{1}$ and $9^{+}_{1}$ states in $^{208}$Tl are calculated to have high excitation energies, and they are dominated by $\pi(0h_{11/2})^{-1}\nu(0j_{15/2})^{1}$ configuration. They can decay through E1 or M1 transitions to other states and are not isomeric states. The $9^{-}_{1}$ state (not drawn in FIG.~\ref{Tleven}) with $\pi(0h_{11/2})^{-1}\nu(0g_{9/2})^{1}$ configuration is predicted to be an isomeric state, which is located $95$ keV above the $7^{+}_{1}$. Although the $12^{+}_{1}$ state with $\pi(0h_{11/2})^{-1}\nu(0j_{15/2})^{1}$ configuration is predicted below the $11^{+}_{1}$ state, it can decay to the $10^{-}_{1,2}$ states through M2 transition with an estimated half-life around $10$ ns.

The situation is different in $^{210,212}$Tl. Because of more valence neutrons, the $11^{+}_{1}$ and $9^{+}_{1}$ states in $^{210,212}$Tl are dominated by $\pi(2s_{1/2})^{-1}\nu(1g_{9/2})^{3}$ configuration and located much lower than those in $^{208}$Tl. The $11^{+}_{1}$ states in $^{210,212}$Tl are predicted to be isomeric in the present calculations. A possible high-spin $\beta$ decaying isomer is suggested to be $11^{+}_{1}$ state in $^{210}$Tl~\cite{Broda2018}, which will be discussed in Sec.~\ref{sec:level6}. The $9^{-}_{1}$ states (not drawn in FIG.~\ref{Tleven}) in $^{210,212}$Tl can decay to $9^{+}_{1}$ states and are not predicted to be isomeric.

\subsection{$^{206,208,210}$Hg}

\begin{figure}
\includegraphics[scale=0.30]{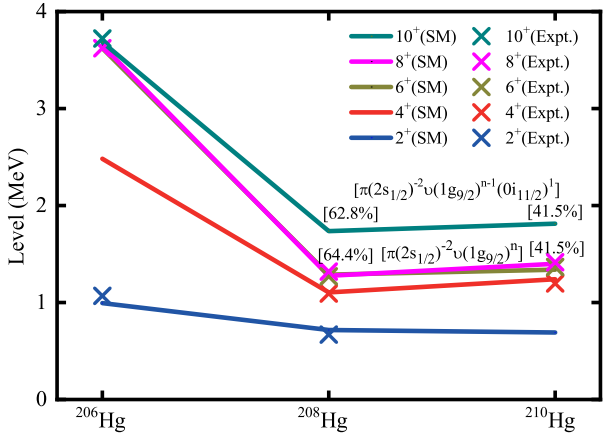}
\caption{\label{Hgeven} (Color online) The calculated and observed levels of $^{206,208,210}$Hg. Observed data are taken from NNDC~\cite{nndc}. The observed $8^{+}_{1}$ states in $^{208,210}$Hg are drawn 0.05 MeV above corresponding $6^{+}_{1}$ states as estimations. Dominant configurations of the $8^{+}_{1}$ and $10^{+}_{1}$ states of $^{208,210}$Hg and their percentages are indicated.}
\end{figure}

From $^{206}$Hg to $^{208}$Hg, the excitation energies of the $6^{+}_{1}$, $8^{+}_{1}$, and $10^{+}_{1}$ states are much reduced, as shown in FIG.~\ref{Hgeven}. Such reduction indicates a transition from proton excitation to neutron excitation. The configurations of the $6^{+}_{1}$, $8^{+}_{1}$, and $10^{+}_{1}$ states in $^{206}$Hg are dominated by two proton hole in the $\pi(0h_{11/2})$ orbit. The $10^{+}_{1}$ state is observed as an isomeric state above the $8^{+}_{1}$ state. But the neutron excitations are dominant in the $6^{+}_{1}$, $8^{+}_{1}$, and $10^{+}_{1}$ states in  $^{208,210}$Hg, which are similar to those of $^{210,212}$Pb. The $8^{+}_{1}$ states of $^{208,210}$Hg were observed as isomeric states consistent with the present calculations.

\subsection{$N=126$ isotones below $^{206}$Hg}

\begin{figure}
\includegraphics[scale=0.30]{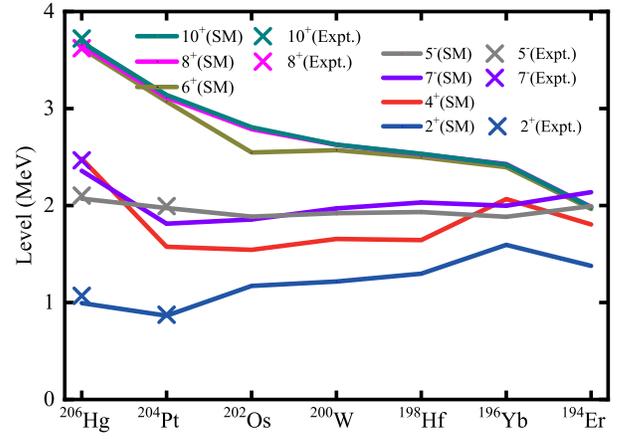}
\caption{\label{126even} (Color online) The calculated and observed levels of $^{206}$Pb, $^{204}$Pt, $^{202}$Os, $^{200}$W, $^{198}$Hf, $^{196}$Yb, and $^{194}$Er. Observed data are taken from Ref.~\cite{Steer2011,Steer2008,Podolyak2009-2}. }
\end{figure}

\begin{figure}
\includegraphics[scale=0.30]{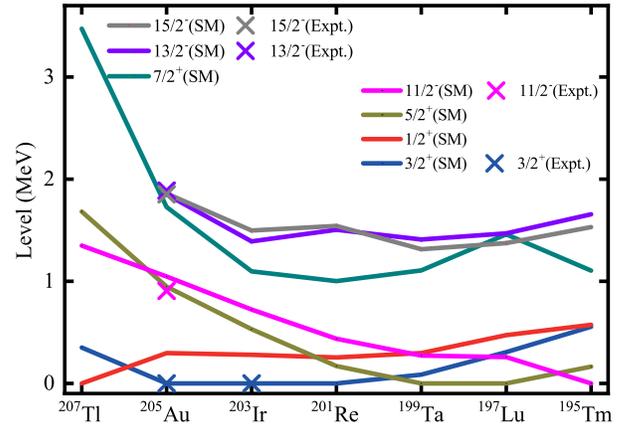}
\caption{\label{126odd} (Color online) The same as FIG. \ref{126even} but for levels of $^{205}$Au, $^{203}$Ir, $^{201}$Re, $^{199}$Ta, $^{197}$Lu, and $^{195}$Tm.}
\end{figure}

\begin{figure}
\includegraphics[scale=0.30]{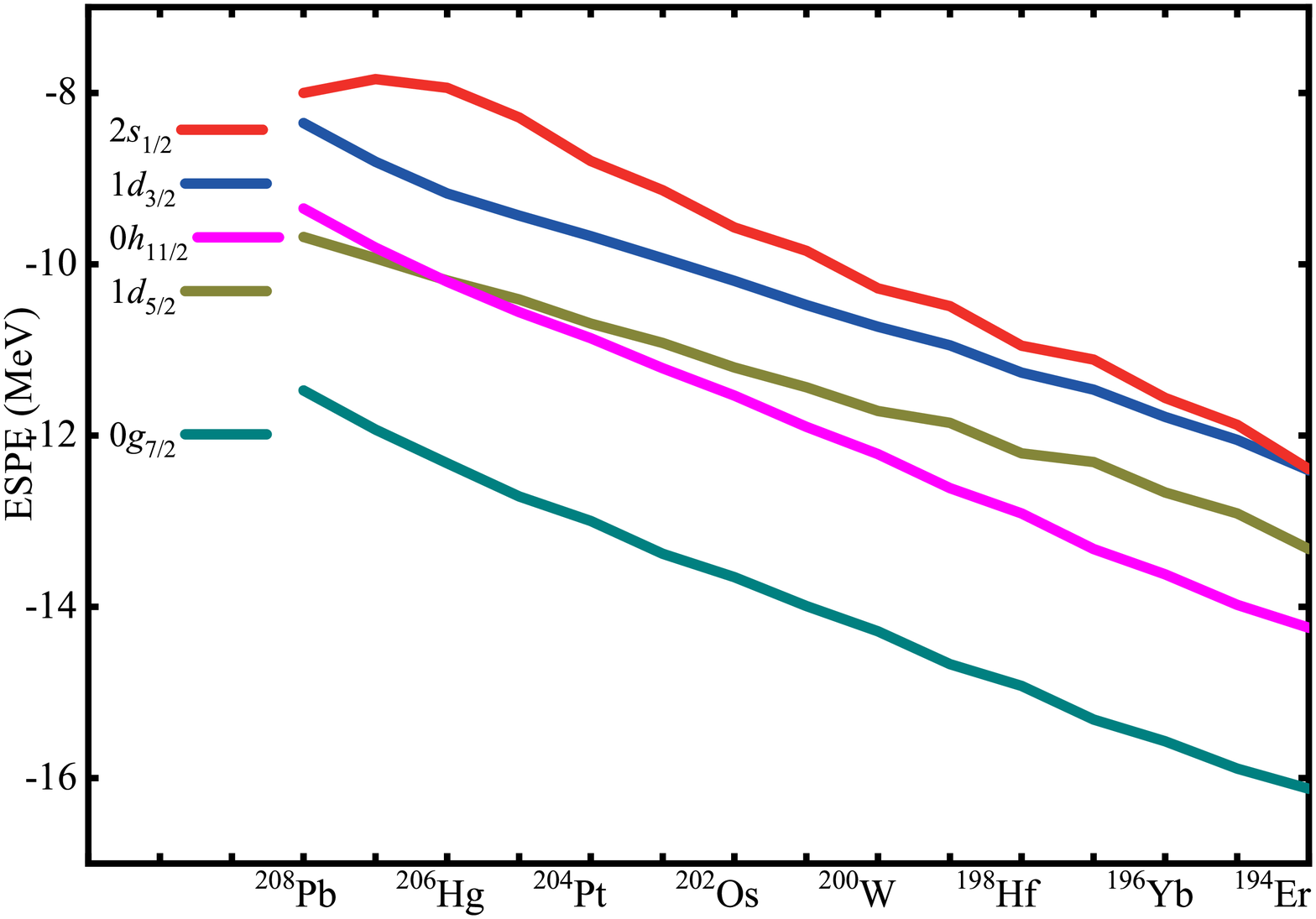}
\caption{\label{ESPEproton} (Color online) Effective single-particle energies of five proton orbits below the $Z=82$ shell closures.}
\end{figure}

There are some recent experimental and theoretical works concerning the spectroscopic properties close to the $N=126$ magic number below Hg isotopes. For example, proton hole states are observed in $^{205}$Au, $^{204}$Pt, and $^{203}$Ir~\cite{Podolyak2009,Podolyak2009-2,Steer2008,Steer2011}. Recent observation of the ground state band of $^{200}$Pt shows that it is a transitional nucleus between the lighter Pt isotopes with $\gamma$-unstable properties and $N = 126$ nucleus $^{204}$Pt with spherical shape~\cite{John2017}. Low-lying states in proton-hole isotopes of Ir, Pt, Au, Hg, and Tl in the $^{208}$Pb region, are investigated through nucleon pair approximation, which agrees well with the observed data~\cite{Jiang2011-1,Jiang2011-2}.

The present work concentrates on the levels of neutron-rich Pb, Tl, and Hg isotopes, especially the possible isomeric states. It can also be used to investigate levels of other nuclei in this region, such as the $N=126$ isotones below Hg isotopes. The results from the present Hamiltonian are presented in FIG. \ref{126even} for even-even nuclei and in FIG. \ref{126odd} for odd-even nuclei. On the whole, the present Hamiltonian generally reproduces well the excited states in the $N=126$ isotones above Ir isotopes. More spectroscopic studies, especially on the experimentally interesting nuclei $^{204}$Pt, $^{202}$Os, $^{200}$W, are necessary to verify the Hamiltonian.

The configurations and excitation energies of the yrast states along $N=126$ isotones are very sensitive to the interaction related to proton $1d_{3/2}$ orbit, which changes the gap between $1d_{3/2}$ and $2s_{1/2}$ orbits. As proton effective single-particle energies (ESPE) presented in FIG. \ref{ESPEproton}, with the modified proton-proton interaction, the $1d_{3/2}$ orbit becomes closer to the $2s_{1/2}$ orbit when the proton number decreases from $^{206}$Hg.

Therefore, the present work provides a possible modification of Hamiltonian, based on the available data of Au, Pt, and Ir isotopes. More data around and below Ir isotopes will be needed for reliable modification of the proton-proton interaction.

\begin{figure}
\includegraphics[scale=0.3]{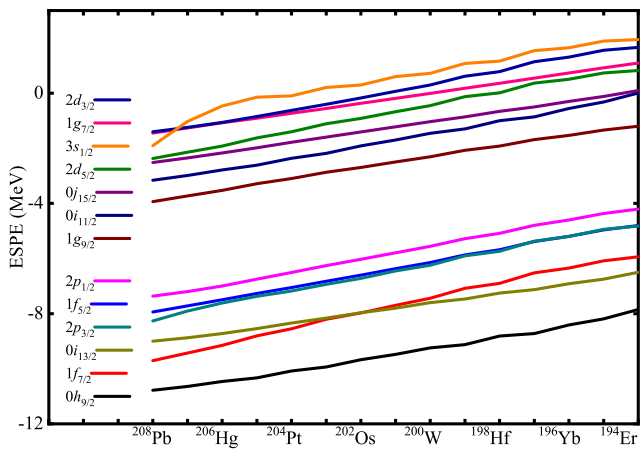}
\caption{\label{ESPE} (Color online) Effective single-particle energies of six (seven) neutron orbits below (beyond) the $N=126$ shell closures.}
\end{figure}

\section{\label{sec:level5}The $N=126$ Shell Gap and Cross-shell Excitations}

Based on the present Hamiltonian, the $N=126$ shell gap and neutron core-excited states are discussed. Figure \ref{ESPE} presents the neutron ESPE along $N=126$ isotones, which are calculated through the present Hamiltonian with the shell-model occupancies. When the proton number decreases, the $N=126$ gap keeps nearly constant with little quenching, and the relative positions of neutron orbits do not change much. Only neutron $1f_{7/2}$ and $0i_{13/2}$ orbits change their orders when proton number changes from 81 to 68.

In light and medium mass region, the orders of some orbits dramatically change along isotopic or isotonic chains, leading to the disappearances of some traditional magic numbers and appearances of some new magic numbers. In the present heavy mass region, two possible mechanisms may be responsible for the little variation of the (sub)shell gaps. One reason is that the effective nucleon-nucleon interaction is relatively weaker in heavy nuclei because of the larger radius. Thus the effect on the (sub)shell gaps is generally less pronounced when a nucleon is removed or added.

\begin{table}
\caption{\label{core_Pb} The excitation energies of observed and possible corresponding calculated states in $^{208}$Pb (unit in MeV). Configurations of the shell-model wave functions are also shown. The experimental data are taken from Ref.~\cite{nndc}.}
\begin{ruledtabular}
\begin{tabular}{ccccc}
      state           &E$_{expt.}$&E$_{SM}$    &configuration  & percentage \\ \hline
      $4^{-}_{1}$     &   3.475  &    3.507   &$\nu(2p_{1/2})^{-1}(1g_{9/2})^{1}$   & 98.4\%\\
      $5^{-}_{1}$     &   3.197  &    3.349   &$\nu(2p_{1/2})^{-1}(1g_{9/2})^{1}$   & 98.1\%\\ \hline
      $2^{-}_{1}$     &   4.229  &    4.241   &$\nu(1f_{5/2})^{-1}(1g_{9/2})^{1}$   & 89.9\%\\
      $3^{-}_{2}$     &   4.051  &    4.115   &$\nu(1f_{5/2})^{-1}(1g_{9/2})^{1}$   & 70.8\%\\
      $4^{-}_{2}$     &   3.946  &    4.048   &$\nu(1f_{5/2})^{-1}(1g_{9/2})^{1}$   & 95.9\%\\
      $5^{-}_{2}$     &   3.708  &    4.012   &$\nu(1f_{5/2})^{-1}(1g_{9/2})^{1}$   & 87.8\%\\
      $6^{-}_{1}$     &   3.919  &    3.995   &$\nu(1f_{5/2})^{-1}(1g_{9/2})^{1}$   & 98.9\%\\
      $7^{-}_{1}$     &   4.037  &    3.833   &$\nu(1f_{5/2})^{-1}(1g_{9/2})^{1}$   & 98.0\%\\ \hline
      $5^{-}_{3}$     &   3.961  &    4.168   &$\nu(2p_{1/2})^{-1}(0i_{11/2})^{1}$  & 92.2\%\\
      $6^{-}_{2}$     &   4.206  &    4.225   &$\nu(2p_{1/2})^{-1}(0i_{11/2})^{1}$  & 98.7\%\\ \hline
      $3^{-}_{3}$     &   4.254  &    4.152   &$\nu(2p_{3/2})^{-1}(1g_{9/2})^{1}$   & 70.9\%\\
      $4^{-}_{3}$     &   3.995  &    4.355   &$\nu(2p_{3/2})^{-1}(1g_{9/2})^{1}$   & 96.4\%\\
      $5^{-}_{4}$     &   4.125  &    4.285   &$\nu(2p_{3/2})^{-1}(1g_{9/2})^{1}$   & 95.1\%\\
      $6^{-}_{3}$     &   4.383  &    4.225   &$\nu(2p_{3/2})^{-1}(1g_{9/2})^{1}$   & 97.3\%\\ \hline
      $ 7^{+}_{1}$    &   4.867  &    4.891   &$\nu(2p_{1/2})^{-1}(0j_{15/2})^{1}$  & 90.2\%\\
      $ 8^{+}_{1}$    &   4.610  &    4.778   &$\nu(2p_{1/2})^{-1}(0j_{15/2})^{1}$  & 95.0\%\\ \hline
      $2^{+}_{2}$     &   4.928  &    4.980   &$\nu(0i_{13/2})^{-1}(1g_{9/2})^{1}$  & 99.5\%\\
      $(3)^{+}_{1}$   &   5.317  &    5.024   &$\nu(0i_{13/2})^{-1}(1g_{9/2})^{1}$  & 99.7\%\\
      $4^{+}_{2}$     &   5.216  &    4.981   &$\nu(0i_{13/2})^{-1}(1g_{9/2})^{1}$  & 99.8\%\\
      $5^{+}_{1}$     &   5.193  &    5.032   &$\nu(0i_{13/2})^{-1}(1g_{9/2})^{1}$  & 97.8\%\\
      $6^{+}_{2}$     &   5.213  &    4.999   &$\nu(0i_{13/2})^{-1}(1g_{9/2})^{1}$  & 98.5\%\\
      $7^{+}_{2}$     &   5.195  &    5.061   &$\nu(0i_{13/2})^{-1}(1g_{9/2})^{1}$  & 89.4\%\\
      $8^{+}_{2}$     &   4.860  &    5.027   &$\nu(0i_{13/2})^{-1}(1g_{9/2})^{1}$  & 95.5\%\\
      $9^{+}_{1}$     &   5.010  &    5.054   &$\nu(0i_{13/2})^{-1}(1g_{9/2})^{1}$  & 98.6\%\\
      $10^{+}_{1}$    &   4.895  &    5.034   &$\nu(0i_{13/2})^{-1}(1g_{9/2})^{1}$  & 91.0\%\\
      $(11^{+}_{1})$  &   5.235  &    5.087   &$\nu(0i_{13/2})^{-1}(1g_{9/2})^{1}$  & 99.7\%\\ \hline
      RMS             &          &    0.169   &                                     &       \\
\end{tabular}
\end{ruledtabular}
\end{table}

The other reason is the configuration mixing. The protons are firstly removed from the $2s_{1/2}$ orbit when proton number changes from 82 to 80. When the proton number is further decreased, protons in $1d_{3/2}$, $1d_{5/2}$, and $0h_{11/2}$ orbits are almost simultaneously removed. Considering the shell evolution mechanism from the tensor force, $j=l+1/2$ and $j=l-1/2$ orbits have opposite contributions to ESPE~\cite{otsuka2001,otsuka2005,otsuka2010}. For example, if $1d_{3/2}$ and $1d_{5/2}$ are fully occupied or fully removed, their tensor interactions with other orbits are almost canceled and have no contributions to shell evolution. It is true that protons in $1d_{3/2}$ orbit are removed more quickly than those in $1d_{5/2}$ and $0h_{11/2}$ orbits, the latter two orbits are both $j=l+1/2$ type orbits. However, because of the configuration mixing, the protons are removed from several orbits, which results in a cancellation of the contributions from the tensor interaction.

Based on the experimental data available in In and Ag isotopes, recent investigations on the $9/2^{+}_{1}$ and $1/2^{-}_{1}$ states in $^{123,125}$Ag~\cite{Chen2019} and $^{101}$In~\cite{Xu2019} show that the evolutions of $Z=40$ subshell are rather slow in Ag isotopes towards $^{132}$Sn and in In isotopes towards $^{100}$Sn, respectively. Shell-model calculations can well describe the shell evolution and spectroscopic properties in Ag and In isotopes through V$_{\text{MU}}$+LS interaction with slight modifications on the strength of the central force~\cite{Chen2019,Xu2019}.

\begin{figure}
\includegraphics[scale=0.24]{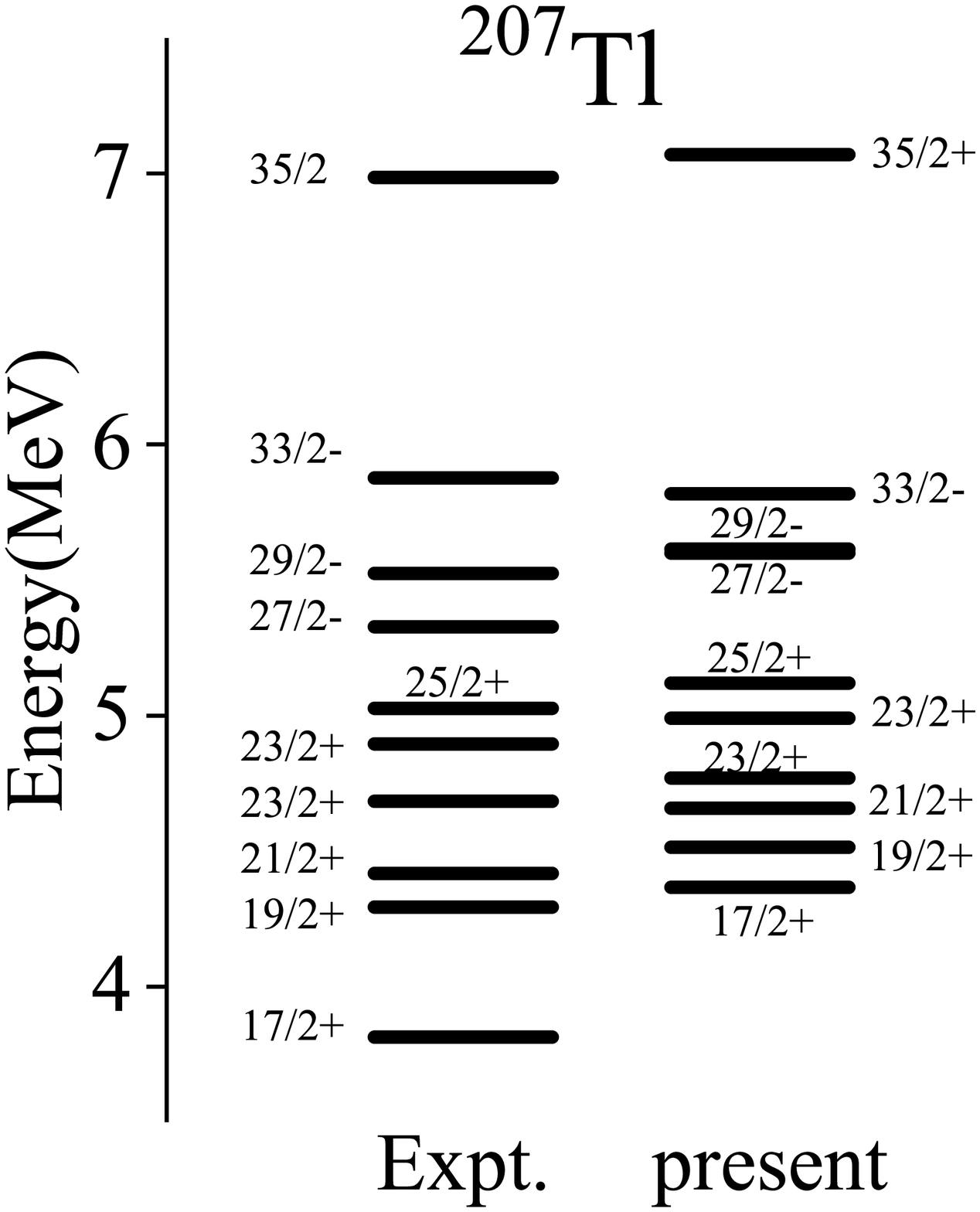}
\includegraphics[scale=0.24]{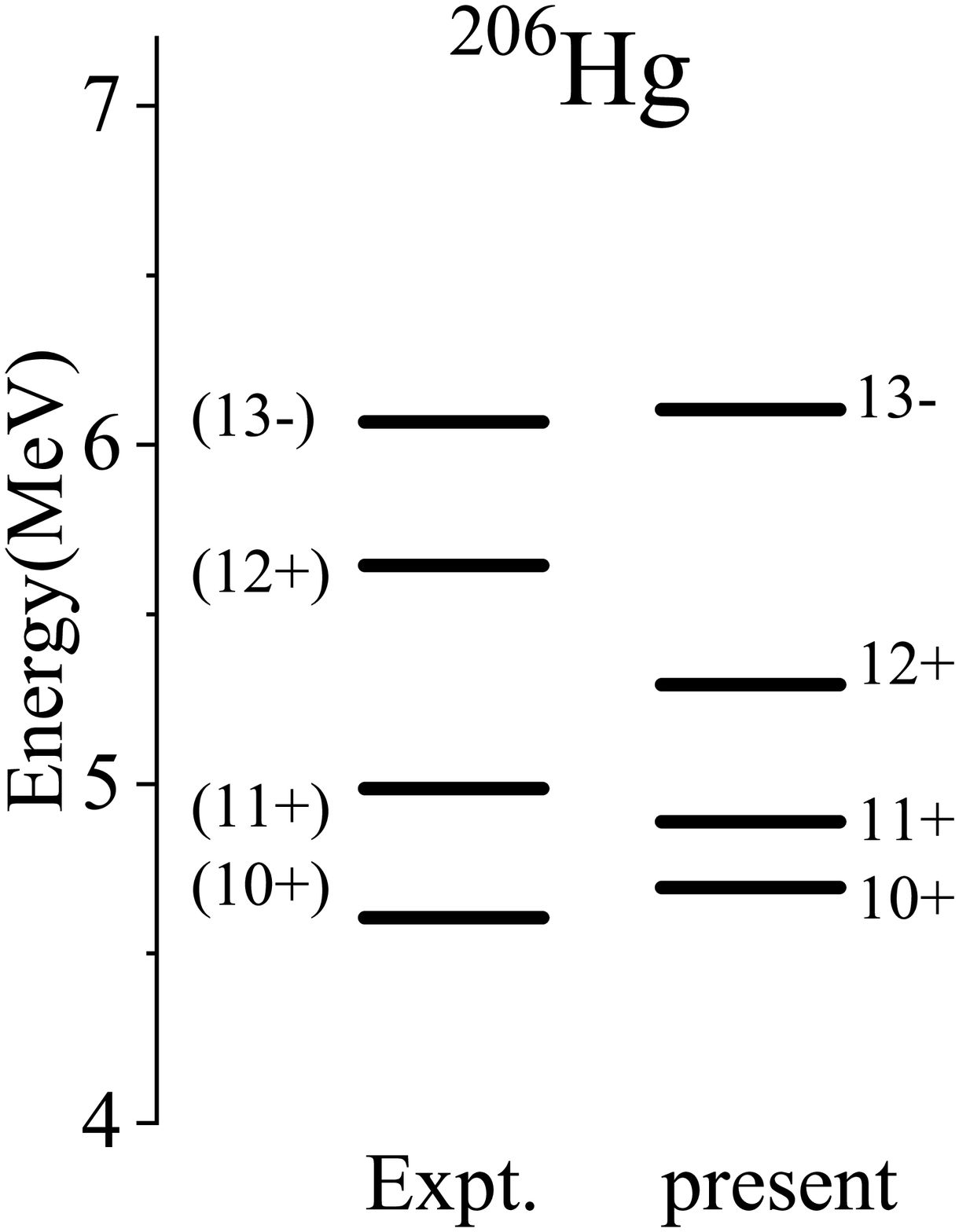}
\caption{\label{corefig} (Color online) The calculated and observed core-excited states in $^{207}$Tl and $^{206}$Hg. Observed data are taken from Ref.~\cite{Wilson2015,nndc}.}
\end{figure}

\begin{table}
\caption{\label{core} The configuration of the core-excited state in $^{207}$Tl and $^{206}$Hg.}
\begin{ruledtabular}
\begin{tabular}{cccc}
      & state   & configuration  & percentage \\ \hline
$^{207}$Tl & $17/2^{+}_{1}$ & $\pi(0h_{11/2})^{-1}\nu(2p_{1/2})^{-1}(1g_{9/2})^{1}$ & 82.4\%\\
      & $19/2^{+}_{1}$ & $\pi(0h_{11/2})^{-1}\nu(2p_{1/2})^{-1}(1g_{9/2})^{1}$ & 82.4\%\\
      & $21/2^{+}_{1}$ & $\pi(2s_{1/2})^{-1}\nu(0i_{13/2})^{-1}(1g_{9/2})^{1}$ & 71.7\%\\
      & $23/2^{+}_{1}$ & $\pi(2s_{1/2})^{-1}\nu(0i_{13/2})^{-1}(1g_{9/2})^{1}$ & 79.9\%\\
      & $23/2^{+}_{2}$ & $\pi(0h_{11/2})^{-1}\nu(1f_{5/2})^{-1}(1g_{9/2})^{1}$ & 81.8\%\\
      &                & $\pi(1d_{3/2})^{-1}\nu(0i_{13/2})^{-1}(1g_{9/2})^{1}$ & 10.3\%\\
      & $25/2^{+}_{1}$ & $\pi(0h_{11/2})^{-1}\nu(1f_{5/2})^{-1}(1g_{9/2})^{1}$ & 81.4\%\\
      & $27/2^{-}_{1}$ & $\pi(0h_{11/2})^{-1}\nu(0i_{13/2})^{-1}(1g_{9/2})^{1}$ & 98.8\%\\
      & $29/2^{-}_{1}$ & $\pi(0h_{11/2})^{-1}\nu(0i_{13/2})^{-1}(1g_{9/2})^{1}$ & 98.6\%\\
      & $33/2^{-}_{1}$ & $\pi(0h_{11/2})^{-1}\nu(0i_{13/2})^{-1}(1g_{9/2})^{1}$ & 99.5\%\\
      & $35/2^{+}_{1}$ & $\pi(0h_{11/2})^{-1}\nu(0i_{13/2})^{-1}(0j_{15/2})^{1}$ & 100.0\%\\
$^{206}$Hg & $10^{+}_{2}$ & $\pi(2s_{1/2})^{-2}\nu(0i_{13/2})^{-1}(1g_{9/2})^{1}$& 47.7\%\\
      &              & $\pi(1d_{3/2})^{-2}\nu(0i_{13/2})^{-1}(1g_{9/2})^{1}$& 9.4\%\\
      &              & $\pi(0h_{11/2})^{-2}\nu(0i_{13/2})^{-1}(1g_{9/2})^{1}$& 6.7\%\\
      & $11^{+}_{1}$ & $\pi(2s_{1/2})^{-2}\nu(0i_{13/2})^{-1}(1g_{9/2})^{1}$& 42.3\%\\
      &              & $\pi(1d_{3/2})^{-2}\nu(0i_{13/2})^{-1}(1g_{9/2})^{1}$& 14.6\%\\
      &              & $\pi(0h_{11/2})^{-2}\nu(0i_{13/2})^{-1}(1g_{9/2})^{1}$& 12.5\%\\
      & $12^{+}_{1}$ & $\pi(2s_{1/2})^{-2}\nu(0i_{13/2})^{-1}(0i_{11/2})^{1}$& 53.8\%\\
      &              & $\pi(1d_{3/2})^{-2}\nu(0i_{13/2})^{-1}(0i_{11/2})^{1}$& 14.6\%\\
      &              & $\pi(0h_{11/2})^{-2}\nu(0i_{13/2})^{-1}(0i_{11/2})^{1}$& 9.3\%\\
      & $13^{-}_{1}$ & $\pi(2s_{1/2})^{-2}\nu(0i_{13/2})^{-1}(0j_{15/2})^{1}$& 37.8\%\\
      &              & $\pi(1d_{3/2})^{-2}\nu(0i_{13/2})^{-1}(0j_{15/2})^{1}$& 14.3\%\\
      &              & $\pi(0h_{11/2})^{-2}\nu(0i_{13/2})^{-1}(0j_{15/2})^{1}$& 11.2\%\\
\end{tabular}
\end{ruledtabular}
\end{table}

\begin{table}
\caption{\label{core_other} The predicted excitation energies of the lowest five one-neutron cross-shell excited states with different spins and parities in $^{205}$Au, $^{204}$Pt, $^{203}$Ir, and $^{202}$Os (unit in MeV).}
\begin{ruledtabular}
\begin{tabular}{cccccc}
  nuclide  & state      &E$_{x}$&  nuclide  & state      &E$_{x}$\\ \hline
$^{205}$Au & $ 9/2^-$ & 2.837 &$^{203}$Ir & $ 9/2^-$   & 2.726\\
           & $11/2^-$ & 2.840 &           & $ 7/2^-$   & 2.732\\
           & $ 7/2^-$ & 2.970 &           & $11/2^-$   & 2.744\\
           & $13/2^-$ & 2.990 &           & $13/2^-$   & 2.768\\
           & $ 5/2^-$ & 3.196 &           & $ 5/2^-$   & 2.835\\
           & $13/2^+$ & 3.605 &           & $13/2^+$   & 3.178\\
           & $11/2^+$ & 3.621 &           & $11/2^+$   & 3.198\\
           & $15/2^+$ & 3.635 &           & $15/2^+$   & 3.200\\
           & $ 9/2^+$ & 3.664 &           & $17/2^+$   & 3.236\\
           & $17/2^+$ & 3.685 &           & $ 9/2^+$   & 3.239\\
$^{204}$Pt & $   5^-$ & 2.845 &$^{202}$Os & $   5^-$   & 2.706\\
           & $   4^-$ & 2.939 &           & $   4^-$   & 2.795\\
           & $   6^-$ & 3.219 &           & $   6^-$   & 3.107\\
           & $   7^-$ & 3.263 &           & $   7^-$   & 3.162\\
           & $   3^-$ & 3.357 &           & $   3^-$   & 3.237\\
           & $   8^+$ & 4.163 &           & $   8^+$   & 3.982\\
           & $   7^+$ & 4.192 &           & $   7^+$   & 4.029\\
           & $   9^+$ & 4.252 &           & $   9^+$   & 4.152\\
           & $   6^+$ & 4.253 &           & $   6^+$   & 4.161\\
           & $   5^+$ & 4.317 &           & $  10^+$   & 4.233\\
\end{tabular}
\end{ruledtabular}
\end{table}

Besides excitations inside one major shell, the observed neutron core-excited states are also well described by the present Hamiltonian. It is seen in Table~\ref{core_Pb} that the excitation energies of low lying states in $^{208}$Pb can be well reproduced by the calculations except for those of the $3^{-}_{1}$, $2^{+}_{1}$, $4^{+}_{1}$, and $6^{+}_{1}$ states.

The results have two indications. On the one hand, the states listed in Table~\ref{core_Pb} are likely to be dominated by one-neutron cross-shell configurations because all theoretically predicted multiplets correspond to the observed states with similar energy. The RMS deviation between observed data and calculated results is as small as $0.169$ MeV. On the other hand, the observed $3^{-}_{1}$, $2^{+}_{1}$, $4^{+}_{1}$, and $6^{+}_{1}$ states are likely to be dominated by more complicated configurations and do not appear in the results of one-neutron cross-shell calculations. Actually, excitation energies of the observed $3^{-}_{2}$, $2^{+}_{2}$, $4^{+}_{2}$, and $6^{+}_{2}$ states are well reproduced by those of the calculated $3^{-}_{1}$, $2^{+}_{1}$, $4^{+}_{1}$, and $6^{+}_{1}$ states.

Such indications can be illustrated by comparing the energy levels and the shell gaps. It can be deduced with the observed binding energies of $^{208, 209}$Pb and $^{209}$Bi \cite{AME2020} that the gap between $N=126$ ($Z=82$) and higher orbits are larger than $3.430$ MeV ($4.204$ MeV). Therefore, the states lying below $4.204$ MeV are probably more dominated by one-neutron cross-shell excitations. The model space truncated in the present work is reasonable for those states. On the other hand, the proton excitation configuration may also contribute to higher levels, contributing to the deviations of the present work. Ref. \cite{Rejmund1999} suggested certain proton-dominant states of $^{208}$Pb by applying single-particle single-hole configurations to reconstruct the wavefunctions but including fewer neutron cross-shell configurations than the present work. It should be noted that neutrons are easier to cross the major shell in the south region of $^{208}$Pb, which is focused on in this work.

The observed $3^{-}_{1}$ state in $^{208}$Pb is well known for its collective octupole character with strong configuration mixing \cite{lane1960,carter1960}, which may be the reason that the percentages of dominating configuration in calculated $3^{-}_{1}$ and $3^{-}_{2}$ states is smaller than those in other states. The present model space is not large enough for the description of the observed $3^{-}_{1}$ state.

Figure~\ref{corefig} shows that the core-excited states of $^{207}$Tl and $^{206}$Hg are well reproduced by the calculations with one-neutron cross-shell excitation. This indicates that the cross-shell interaction is reasonable up to $Z=80$. The $17/2^{+}$ state in $^{207}$Tl is also a collective octupole state with $\pi(0h_{11/2})^{-1}$ coupled to the $3^{-}_{1}$ state in $^{208}$Pb, which is the reason that the excitation energy of the calculated state with simple configuration has large discrepancy with the experimental data.

Detailed configurations of these states are listed in Table \ref{core}. Although the ground state of $^{207}$Tl is dominated by a proton hole in the $2s_{1/2}$ orbit, many core-excited states have a proton hole in the $0h_{11/2}$ orbit, which contributes to the high angular momenta of these states.

Comparing with the shell-model calculations with KHH7B interaction~\cite{Wilson2015}, the present results give quite similar configurations for core-excited states (except for the $23/2^{+}_{2}$ state) in $^{207}$Tl. For $23/2^{+}_{2}$ state in $^{207}$Tl, the present work gives a dominant $\pi(0h_{11/2})^{-1}\nu(1f_{5/2})^{-1}(1g_{9/2})^{1}$ configuration, while KHH7B interaction gives a strong configuration mixing between $\pi(0h_{11/2})^{-1}\nu(1f_{5/2})^{-1}(1g_{9/2})^{1}$ and $\pi(1d_{3/2})^{-1}\nu(0i_{13/2})^{-1}(1g_{9/2})^{1}$ configurations~\cite{Wilson2015}.

Contrary to $^{207}$Tl, the angular momenta of observed core-excited states in $^{206}$Hg are mostly originated from the coupling of neutrons. Two proton holes in the $2s_{1/2}$ orbit are dominant proton configuration and have no contribution to the angular momenta of these states. Besides the core-excited states in nuclei discussed above, $^{208}$Pb, $^{207}$Tl, and $^{206}$Hg, the $1/2^{-}_{1}$ state in $^{209}$Pb is well reproduced by a state with $65.6\%$ $\nu(2p_{1/2})^{-1}(1g_{9/2})^{2}$, $18.5\%$ $\nu(2p_{1/2})^{-1}(0i_{11/2})^{2}$, and, $9.6\%$ $\nu(2p_{1/2})^{-1}(0i_{11/2})^{2}$ configurations. The observed and calculated excitation energies are $2.149$ and $2.219$ MeV, respectively.

Some predictions are shown in Table~\ref{core_other} for the excitation energies of the five lowest one-neutron cross-shell excited states with different spins and parities in $^{205}$Au, $^{204}$Pt, $^{203}$Ir, and $^{202}$Os. The spin-parity of the calculated lowest states are quite similar between $^{204}$Pt and $^{202}$Os and between $^{205}$Au and $^{203}$Ir, respectively. But it becomes more difficult to find such cross-shell excited states when the proton number decreases. With more proton-holes, normal states without cross-shell excitations are easier to have different spins and parities, while cross-shell excited states have fewer probabilities of being observed as yrast states.

\section{\label{sec:level6}Electric Quadrupole Properties}

\begin{table*}
\caption{\label{E2-1} The calculated and observed electric quadrupole moments Q (for the first 26 rows) or reduced transition strength B(E2) (for the last 6 rows). Observed data are taken from Ref.~\cite{Stone2016}. M$_{p}$ and M$_{n}$ are proton and neutron E2 transition matrix elements, respectively. SM1 and SM2 are shell-model calculations with two sets of effective charges, $e_{p}=1.59$, $e_{n}=0.85$ and $e_{p}=1.80$, $e_{n}=0.80$, respectively. The units for E$_{x}$, Q, B(E2), and M are keV, eb, e$^{2}$b$^{2}$, and b, respectively. RMS deviation is calculated from Q and $\sqrt{\text(B(E2))}$.}
\begin{ruledtabular}
\begin{tabular}{cccccccc}
nuclides & E$_{x}$ &J$^{\pi}$& M$_{p}$& M$_{n}$& Q$_{SM1}$ or ${B(E2)}$$_{SM1}$& Q$_{SM2}$ or ${B(E2)}$$_{SM2}$& Q$_{expt.}$ or ${B(E2)}$$_{expt.}$ \\ \hline
$^{211}$Pb&0     &  9/2$^+$                               &   0.000  & -0.217   & -0.185   &-0.174  &    +0.09(6)     \\
$^{209}$Pb&0     &  9/2$^+$                               &   0.000  & -0.327   & -0.278   &-0.262  &    -0.27(17)    \\
$^{206}$Pb&803   &    2$^+$                               &   0.000  & 0.327    & 0.278    &0.258   &    +0.05(9)     \\
          &2200  &    7$^-$                               &   0.000  & 0.483    & 0.410    &0.386   &    0.33(5)      \\
          &4027  &   12$^+$                               &   0.000  & 0.551    & 0.469    &0.441   &    [0.51(2)]    \\
$^{205}$Pb&0     &  5/2$^-$                               &   0.000  & 0.264    & 0.224    &0.211   &    +0.23(4)     \\
          &1014  & 13/2$^+$                               &   0.000  & 0.523    & 0.444    &0.418   &    0.30(5)      \\
          &3196  & 25/2$^-$                               &   0.000  & 0.706    & 0.600    &0.565   &    0.63(3)      \\
$^{204}$Pb&899   &    2$^+$                               &   0.000  & -0.043   & -0.036   &-0.034  &    +0.23(9)     \\
          &1274  &    4$^+$                               &   0.000  & 0.347    & 0.295    &0.277   &    0.44(2)      \\
$^{203}$Pb&0     &  5/2$^-$                               &   0.000  & 0.100    & 0.085    &0.080   &    +0.10(5)     \\
          &1921  & 21/2$^+$                               &   0.000  & 0.909    & 0.773    &0.727   &    0.85(3)      \\
$^{202}$Pb&2170  &    9$^-$                               &   0.000  & 0.638    & 0.542    &0.510   &    +0.58(9)     \\
          &2208  &    7$^-$                               &   0.000  & 0.369    & 0.314    &0.295   &    0.28(2)      \\
$^{201}$Pb&0     &  5/2$^-$                               &   0.000  & -0.042   & -0.035   &-0.033  &    0.01(4)      \\
          &2719  & 25/2$^-$                               &   0.000  & 0.512    & 0.435    &0.410   &    0.46(2)      \\
$^{200}$Pb&2154  &    7$^-$                               &   0.000  & 0.414    & 0.352    &0.332   &    0.32(2)      \\
          &2183  &    9$^-$                               &   0.000  & 0.478    & 0.406    &0.382   &    0.40(2)      \\
          &3006  &   12$^+$                               &   0.000  & 0.589    & 0.501    &0.471   &    0.79(3)      \\
$^{199}$Pb&0     &  3/2$^-$                               &   0.000  & 0.074    & 0.063    &0.059   &    +0.08(9)     \\
$^{206}$Hg&2102  &    5$^-$                               &   0.381  & 0.000    & 0.605    &0.684   &    0.74(15)     \\
$^{204}$Hg&437   &    2$^+$                               &   0.236  & 0.326    & 0.653    &0.686   &    +0.4(2)      \\
$^{203}$Hg&0     &  5/2$^-$                               &   0.137  & 0.301    & 0.474    &0.488   &    +0.344(7)    \\
$^{202}$Hg&440   &    2$^+$                               &   0.232  & 0.396    & 0.706    &0.735   &    +0.87(13)    \\
$^{200}$Hg&368   &    2$^+$                               &   0.236  & 0.492    & 0.793    &0.818   &    +0.96(11)    \\
$^{199}$Hg&532   & 13/2$^+$                               &   0.285  & 0.963    & 1.272    &1.283   &    +1.2(3)      \\
$^{210}$Pb&800   & B(E2;0$^{+}$ $\rightarrow$ $2^{+}$)    &   0.000  & 0.284    & 0.058    &0.052   &    0.051(15)    \\
$^{206}$Pb&803   & B(E2;0$^{+}$ $\rightarrow$ $2^{+}$)    &   0.000  & 0.373    & 0.100    &0.089   &    0.0989(28)   \\
$^{204}$Pb&899   & B(E2;0$^{+}$ $\rightarrow$ $2^{+}$)    &   0.000  & 0.470    & 0.160    &0.142   &    0.1587(69)   \\
$^{204}$Hg&437   & B(E2;0$^{+}$ $\rightarrow$ $2^{+}$)    &   0.258  & 0.357    & 0.509    &0.562   &    0.424(21)    \\
$^{202}$Hg&440   & B(E2;0$^{+}$ $\rightarrow$ $2^{+}$)    &   0.263  & 0.499    & 0.709    &0.761   &    0.615(21)    \\
$^{200}$Hg&368   & B(E2;0$^{+}$ $\rightarrow$ $2^{+}$)    &   0.265  & 0.611    & 0.884    &0.931   &    0.855(28)    \\
RMS       &      &          &                 &           & 0.127    &0.129 &            \\
\end{tabular}
\end{ruledtabular}
\end{table*}

Generally speaking, effective charges are larger with a heavier core in shell-model calculations, because a heavier core normally has stronger core polarization effect. For example, proton (neutron) effective charges are around $1.25$ ($0.25$) for $psd$ shell calculations with $^{4}$He core \cite{yuan2012}, and $1.7$ ($0.7$) for nuclei around $^{132}$Sn \cite{yuan2016}. However, for nuclei around $^{208}$Pb, there are few investigations, especially systematic investigation, on the values of effective charges.

In the present study, $26$ quadrupole moments and $6$ B(E2) values in Pb and Hg isotopes are considered to investigate the values of effective charges. If proton and neutron effective charges $e_{p}$ and $e_{n}$ are constrained as $e_{p}=1+e_{n}$, their values are $1.80$ and $0.80$ from fitting to the $32$ observed data. If the fitting is free without any constraints, $e_{p}$ and $e_{n}$ are $1.59$ and $0.85$, respectively. The values of effective charges obtained with $e_{p}=1+e_{n}$ relationship are a little larger for the present $^{208}$Pb region compared with those for the $^{132}$Sn region.

\begin{figure}
\includegraphics[scale=0.23]{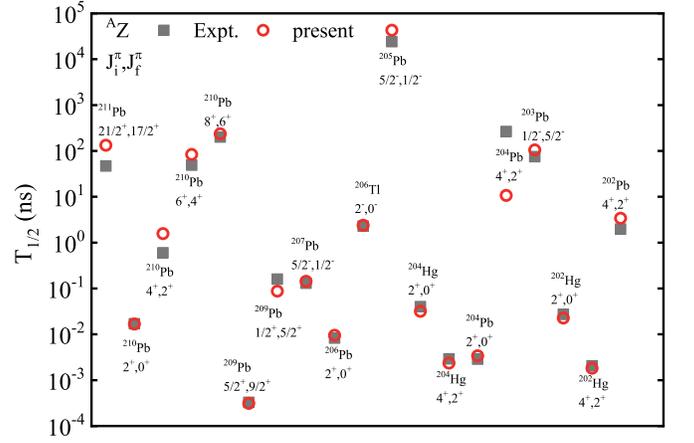}
\caption{\label{E2-2} (Color online) The calculated and observed half-lives corresponding to E2 transitions. The calculated results are based on shell-model transition rates and observed transition energies. Observed data are taken from NNDC~\cite{nndc}.  }
\end{figure}

The two sets of the effective charges obtained do not give large difference in the results for the $32$ observed E2 properties. The detailed results for the quadrupole moments and B(E2) values are presented in Table \ref{E2-1}. Almost all observed data are well reproduced by the shell-model transition matrix elements with the present two sets of effective charges. Although the calculations give the opposite sign for the quadrupole moments of the ground state of $^{211}$Pb and the $2^{+}_{1}$ state in $^{204}$Pb, the absolute deviations from the experimental data are not large. Therefore, the theoretical results presented in FIG. \ref{E2-2}, Table \ref{E2-3}, and Table \ref{E2-4}, are derived with $e_{p}=1.80$ and $e_{n}=0.80$.

The observed quadrupole moment of the ground state ($3/2^{-}$) of $^{201}$Hg is +0.387(6) eb. Shell-model calculations fail to reproduce the $3/2^{-}$ state to be the ground state. The calculations give three very close $3/2^{-}$ states with excitation energies (quadrupole moments) $0.111$ MeV (-0.07 eb), $0.170$ MeV (0.09 eb), and $0.461$ MeV (0.25 eb), respectively, where the quadrupole moments are calculated with $e_{p}=1.80$ and $e_{n}=0.80$. It seems that the third $3/2^{-}$ state is more likely to be the observed ground state because of its quadrupole moment. The magnetic moment of this $3/2^{-}$ state will be discussed in the Sec. \ref{sec:level7}.

\begin{table}
\caption{\label{E2-3} The prediction of possible isomers including transition energy ($\delta$E), B(E2), and half-lives.}
\begin{ruledtabular}
\begin{tabular}{cccccc}
nuclide&J$^{\pi}_{i}$& J$^{\pi}_{f}$& $\delta$E (MeV)  &  B(E2)(e$^{2}$fm$^{4}$)    & half-life ($\mu$s)\\ \hline
$^{215}$Pb &17/2$^+$  & 13/2$^+$  & 0.33  &  28.08    &  0.0047     \\
           &21/2$^+$  & 17/2$^+$  & 0.12  &  31.44    &  0.18     \\
$^{213}$Pb &17/2$^+$  & 13/2$^+$  & 0.29  &  33.46    &  0.0069     \\
           &21/2$^+$  & 17/2$^+$  & 0.12  &  0.45     &  12.13    \\
$^{213}$Tl &13/2$^+$  &  9/2$^+$  & 0.09  &  0.01    &  1812.75  \\
           &17/2$^+$  & 13/2$^+$  & 0.05  &  0.03     &  444.55   \\
$^{212}$Tl &  11$^+$  &    9$^+$  & 0.11  &  2.27     &  2.93     \\
$^{211}$Tl &13/2$^+$  &  9/2$^+$  & 0.09  &  66.27    &  0.14     \\
           &17/2$^+$  & 13/2$^+$  & 0.05  &  18.66    &  0.65     \\
$^{210}$Tl &  11$^+$  &    9$^+$  & 0.11  &  140.76   &  0.053     \\
$^{210}$Hg &   6$^+$  &    4$^+$  & 0.10  &  69.26    &  0.12     \\
           &   8$^+$  &    6$^+$  & 0.06  &  28.03    &  0.45     \\
$^{209}$Hg &17/2$^+$  & 13/2$^+$  & 0.38  &  433.65   &  0.00016     \\
           &21/2$^+$  & 17/2$^+$  & 0.12  &  213.71   &  0.029     \\
\end{tabular}
\end{ruledtabular}
\end{table}

\begin{table}
\caption{\label{E2-4} The estimation of unknown transition energies ($\delta$E) of isomeric states through observed half-lives and calculated B(E2). Observed half-lives are taken from NNDC~\cite{nndc}.}
\begin{ruledtabular}
\begin{tabular}{cccccc}
nuclide&J$^{\pi}_{i}$& J$^{\pi}_{f}$& half-life   &  B(E2)(e$^{2}$fm$^{4}$)    & $\delta$E (MeV)\\ \hline
$^{216}$Pb &   8$^+$  &   6$^+$  & 0.40(4)$\mu$s  &    14.97   &  0.12    \\
$^{214}$Pb &   8$^+$  &   6$^+$  & 6.2(3) $\mu$s  &    0.25    &  0.19    \\
$^{211}$Pb &27/2$^+$  &23/2$^+$  & 159(28) ns     &    66.83   &  0.10  \\
$^{209}$Tl &17/2$^+$  &13/2$^+$  & 95(11) ns	  &    62.73   &  0.12  \\
$^{210}$Hg &   8$^+$  &   6$^+$  & 2(1)  $\mu$s	  &    28.03   &  0.01  \\
$^{208}$Hg &   8$^+$  &   6$^+$  & 99(14) ns      &    90.82   &  0.10    \\
$^{201}$Pt &19/2$^+$  &15/2$^+$  & 21(3) ns	  &  1121.05   &  0.01    \\
\end{tabular}
\end{ruledtabular}
\end{table}

Some observed E2 transition data are presented by half-lives instead of transition strengths. Because of uncertainties of the half-lives, they are not used to deduce the B(E2) values and to fit effective charges. Some half-lives of nuclei in this region are calculated by shell-model transition rates and observed transition energies. It should be noted that electron conversion coefficients need to be considered when transition energies are small. The corresponding electron conversion coefficients are calculated by BrIcc code from ANU \cite{BrIcc}.

It is seen in FIG. \ref{E2-2} that the half-lives are well reproduced, which means that the shell model provides reasonable transition rates. The only exception is the half-life of $^{204}$Pb, of which the observed data is nearly $25$ times longer than the calculation. In other words, the corresponding transition rate is around $25$ times overestimated. As shown in Table \ref{E2-1}, the B(E2;0$^{+}$ $\rightarrow$ 2$^{+}$) value and the quadrupole moments of the $4^{+}$ state in $^{204}$Pb are reasonably reproduced. The B(E2;2$^{+}$ $\rightarrow$ 4$^{+}$) value seems unexpectedly overestimated by the shell-model calculation. But one should note that the absolute B(E2;2$^{+}$ $\rightarrow$ 4$^{+}$) value evaluated from the half-life is only $0.52$ e$^{2}$fm$^{4}$, which is rather small and difficult to be exactly described.

Based on the reliable agreement between the observed and calculated E2 properties, some possible isomeric states, which decay through E2 transition with small decay energies, are presented in Table \ref{E2-3} with effective charges $e_{p}=1.80$ and $e_{n}=0.80$. The half-lives range from several ns to several $\mu$s.

Two isomeric states with rather long half-lives are predicted for $17/2^+$ and $13/2^+$ states in $^{213}$Tl. The $17/2^+$, $13/2^+$, and $9/2^+$ states in $^{213}$Tl are dominated by similar configurations, $\pi(2s_{1/2})^{-1}\nu(1g_{9/2})^{6}$ and $\pi(2s_{1/2})^{-1}\nu(1g_{9/2})^{4}(0i_{11/2})^{2}$. In the two transitions $17/2^{+} \rightarrow 13/2^{+}$ and $13/2^{+} \rightarrow 9/2^{+}$, the one body transition densities involving $\nu(1g_{9/2})$ and $\nu(0i_{11/2})$ orbits are not large and greatly canceled by other components, which leads to small transition matrix elements. The experimental data of $^{213}$Tl are rather limited~\cite{Gottardo2019}, and these two predicted isomers are not found in a recent experiment.

The predicted half-life of the $17/2^+$ state in $^{211}$Tl agrees quite well with the recently observed $0.58(8)$ $\mu$s isomer~\cite{Gottardo2019}. Ref.~\cite{Gottardo2019} also suggested this isomer to be a $17/2^+$ state. Evidence was shown for a high-spin $\beta$ decaying isomer in $^{210}$Tl~\cite{Broda2018}. The present calculation gives a $0.053$ $\mu$s half-life for the $11^{+}_{1}$ state, which is much smaller than the lower limit value of $3$ $\mu$s of the $\beta$ decaying isomer~\cite{Broda2018}. If the $11^{+}_{1}$ state is assumed to be a candidate for the $\beta$ decaying isomer, there are two possibilities. One possibility is that the decay energy from $11^{+}_{1}$ to $9^{+}_{1}$ is very small. For example, the half-life is estimated to be $2.7$ $\mu$s when decay energy is $1.7$ keV. Another possibility is that the $11^{+}_{1}$ state is located below the $9^{+}_{1}$ state and decays to other states with lower spins.

Based on the observed half-lives and calculated E2 properties, it is possible to estimate unknown transition energies of some isomeric states in the south region of Pb, such as transitions from 8$^+_{1}$ to 6$^+_{1}$ in $^{214,216}$Pb and $^{208,210}$Hg~\cite{nndc}. The estimated energies are rather small, which are listed in Table \ref{E2-4}. The observed half-lives are given with fair uncertainties, which induce around $0.01$ MeV uncertainties on the decay energies.  The transition energies in $^{214,216}$Pb are assumed to be between 0.02 and 0.09 MeV~\cite{Gottardo2012}, which are slightly smaller than the present estimations. The present estimations provide reasonable information for future studies.

\begin{table*}
\caption{\label{gfactor} Six sets of effective $g$ factors (unit in $\mu_{n}$) with RMS deviations between observed and calculated magnetic moments of $44$ states of nuclei in the south region of $^{208}$Pb. RMS deviation from free $g$ factors is also presented for comparison. Set1 (Set3) is fitted to all $44$ magnetic moments without (with) considering the $g^{(t)}_{p,n}$ terms. Also without considering $g^{(t)}_{p,n}$ terms, Set4 is fitted separately to $7$ states in nuclei with $N\geq126$ and $37$ states in nuclei with $N<126$. Set2 (Set5) is fitted to the same data set as Set1 (Set4) with considering the $g^{(t)}_{p,n}$ terms while the $g^{(l)}_{p,n}$ and $g^{(s)}_{p.n}$ terms are kept the same as Set1 (Set4). Set6 is fitted in a similar way as fitting Set4 but without consideration of $g^{(l)}_{n}$ term. Uncertainties of the fit parameters are indicated in parentheses.   }
\begin{ruledtabular}
\begin{tabular}{cccccccc}
                 & RMS & $g^{(l)}_{p}$ & $g^{(l)}_{n}$ &$g^{(s)}_{p}$ & $g^{(s)}_{n}$ & $g^{(t)}_{p}$ & $g^{(t)}_{n}$ \\ \hline
Free             & 0.77& 1            & 0           & 5.586         &  -3.826         & 0            &  0            \\
Set1             & 0.21& 1.035(0.027) & 0.015(0.007)& 3.735 (0.165) &  -2.400 (0.123) & 0            &  0            \\
Set2             & 0.20& 1.035(0.027) & 0.015(0.007)& 3.735 (0.165) &  -2.400 (0.123) & 0.455(0.231) & -0.042(0.079) \\
Set3             & 0.19& 1.021(0.027) & 0.005(0.011)& 3.554 (0.173) &  -2.043 (0.332) & 0.646(0.279) & -0.297(0.254) \\
Set4             & 0.13&              &             &               &                 &              &               \\
     ($N\geq126$)&     & 1.030(0.005) &-0.072(0.071)& 3.710 (0.043) &  -2.386 (0.603) & 0            &  0            \\
     ($N<126$)   &     & 1.026(0.022) & 0.007(0.005)& 3.797 (0.126) &  -2.082 (0.090) & 0            &  0            \\
Set5             & 0.12&              &             &               &                 &              &               \\
     ($N\geq126$)&     & 1.030(0.005) &-0.072(0.071)& 3.710 (0.043) &  -2.386 (0.603) & 0.100(0.072) &  0.009(0.018) \\
     ($N<126$)   &     & 1.026(0.022) & 0.007(0.005)& 3.797 (0.126) &  -2.082 (0.090) & 0.238(0.155) & -0.095(0.056) \\
Set6             & 0.13&              &             &               &                 &              &               \\
     ($N\geq126$)&     & 1.030(0.005) & 0           & 3.728 (0.040) &  -3.001 (0.020) & 0            &  0            \\
     ($N<126$)   &     & 1.030(0.022) & 0           & 3.789 (0.127) &  -1.990 (0.058) & 0            &  0            \\
\end{tabular}
\end{ruledtabular}
\end{table*}

\section{\label{sec:level7}magnetic dipole moments}

\begin{figure}
\includegraphics[scale=0.25]{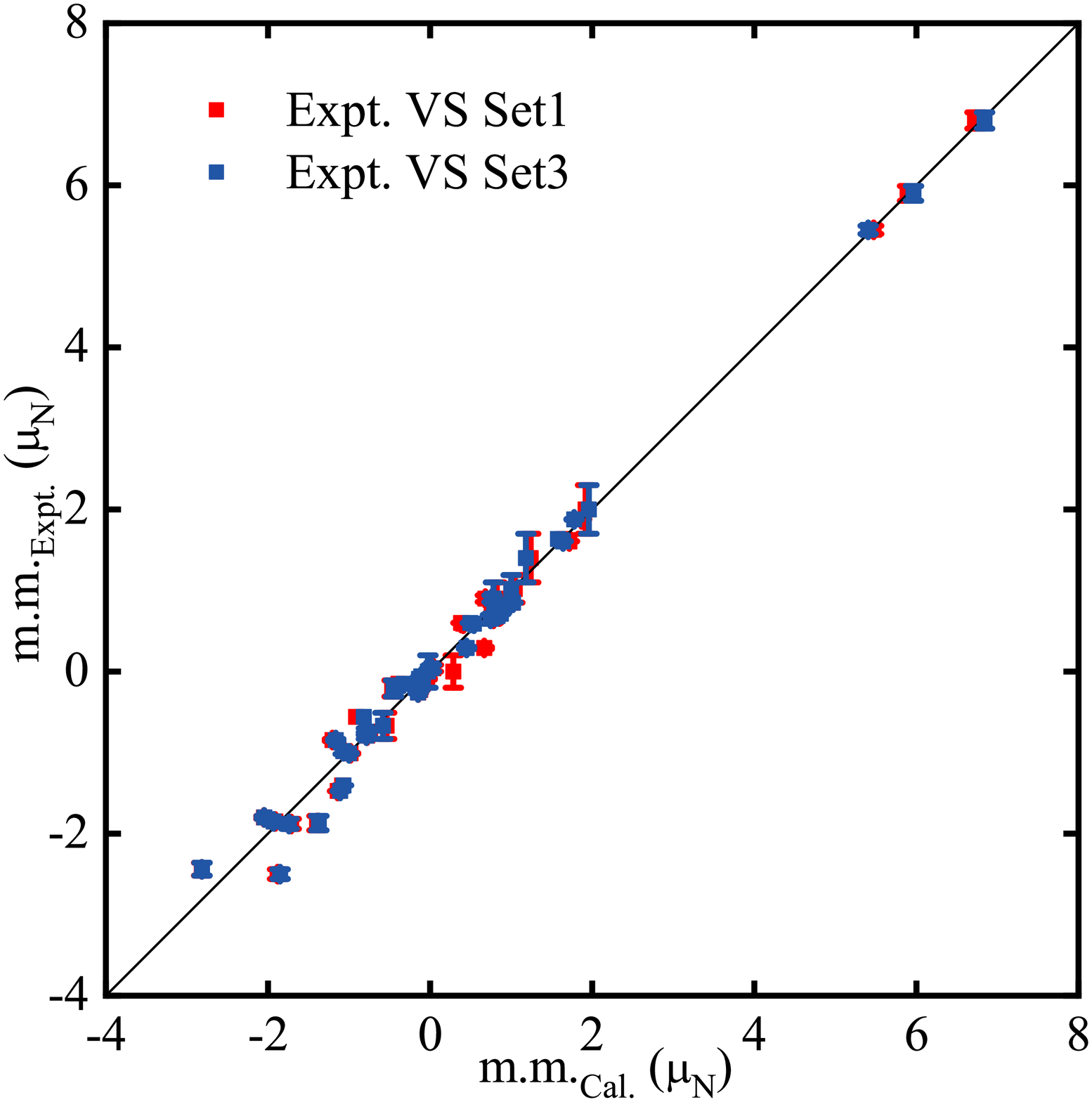}
\includegraphics[scale=0.25]{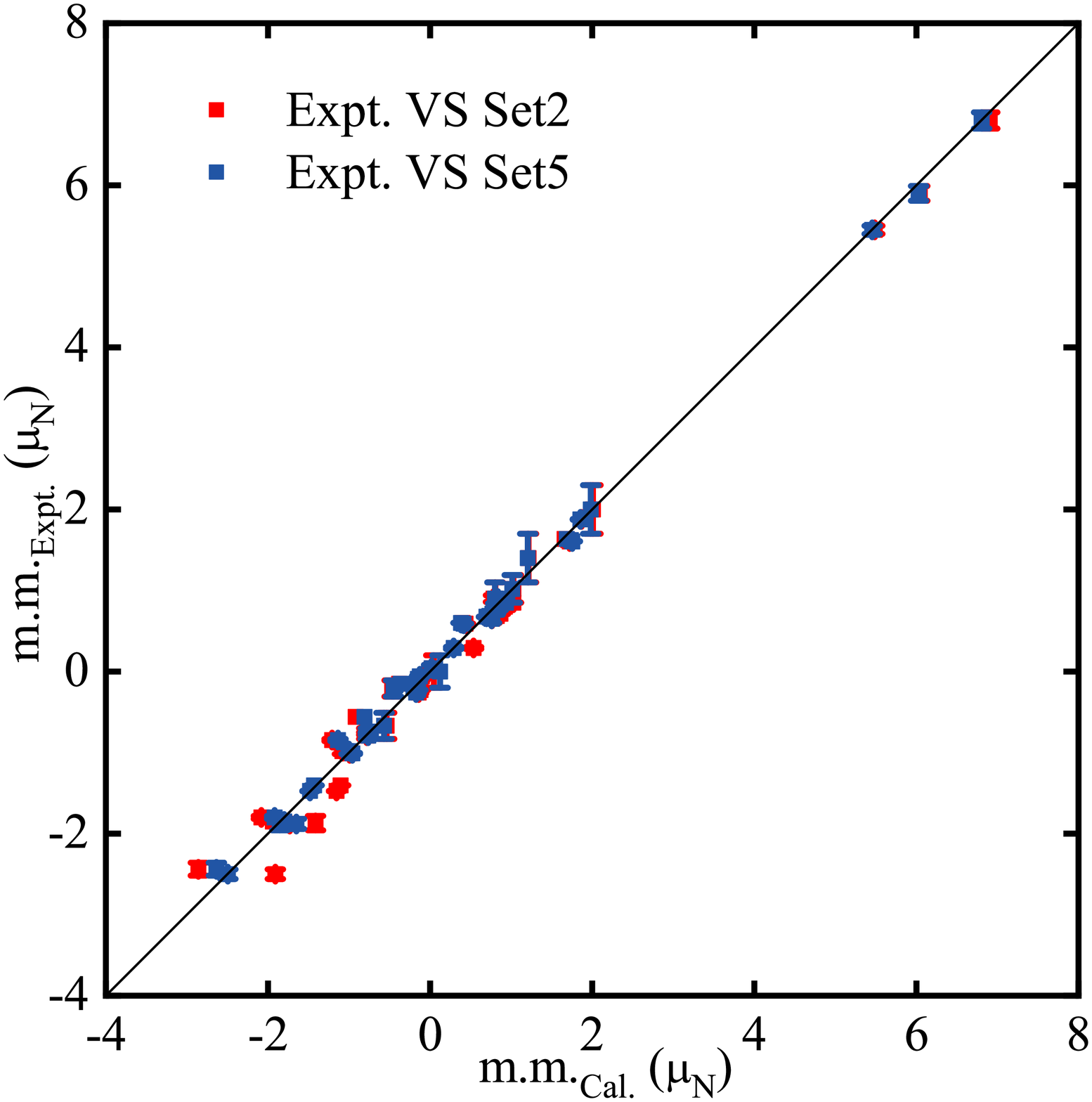}
\caption{\label{mmfig} (Color online) The comparison between calculated and observed magnetic moments. Observed data are taken from Ref.~\cite{Stone2005}. }
\end{figure}

Similar to the investigations on the effective charges, effective $g$ factors are investigated based on $44$ magnetic moments in this region. The $M1$ operator used in the present work includes six terms \cite{Richter2008PRC},
\begin{eqnarray}
  O(M1) &=& \sqrt{\frac{3}{4\pi}}\{\overrightarrow{l}_{p}g^{(l)}_{p} + \overrightarrow{l}_{n}g^{(l)}_{n} + \overrightarrow{s}_{p}g^{(s)}_{p} + \overrightarrow{s}_{n}g^{(s)}_{n}  \nonumber \\
    +  &\sqrt{8\pi}&[(Y^{2}_{p}\otimes\overrightarrow{s}_{p})^{(1)}g^{(t)}_{p}+ (Y^{2}_{n}\otimes\overrightarrow{s}_{n})^{(1)}g^{(t)}_{n})]\},
\end{eqnarray}
where $\overrightarrow{l}$, $\overrightarrow{s}$, and $Y^{2}\otimes\overrightarrow{s}$ are the angular momentum, spin, and tensor terms of the operator, respectively, and $g^{(l)}_{p,n}$, $g^{(s)}_{p,n}$, and $g^{(t)}_{p,n}$ are corresponding proton and neutron $g$ factors. For each term of the operator, the nuclear matrix element is calculated from the single-particle matrix element through the single-particle basis and the one body transition density through the calculated shell-model wave functions. The magnetic moment is calculated through,
\begin{equation}
  \mu = g^{(l)}_{p}l_{p} + g^{(l)}_{n}l_{n} + g^{(s)}_{p}s_{p} + g^{(s)}_{n}s_{n} + g^{(t)}_{p}t_{p} + g^{(t)}_{n}t_{n},
\end{equation}
where $l_{p,n}$, $s_{p,n}$, and $t_{p,n}$ are the proton and neutron angular momentum, spin, and tensor terms of nuclear matrix elements, respectively.

Because shell-model calculations are performed in the truncated model space, effective $g$ factors should be used instead of the free ones. In Table \ref{gfactor}, six sets of effective $g$ factors are presented, which are fitted to the $44$ magnetic moments listed in Table \ref{mm} with the shell-model nuclear matrix elements. Set1 (Set3) of $g$ factors is fitted to all $44$ magnetic moments without (with) considering the $g^{(t)}_{p,n}$ terms. Also without considering the $g^{(t)}_{p,n}$ terms, Set4 of $g$ factors is fitted separately to the $7$ magnetic moments in nuclei with $N\geq126$ and the $37$ magnetic moments in nuclei with $N<126$. Set2 (Set5) of $g$ factors is fitted to the same data set as for Set1 (Set4) by considering the $g^{(t)}_{p,n}$ terms while the $g^{(l)}_{p,n}$ and $g^{(s)}_{p.n}$ terms are kept the same as those in Set1 (Set4). Set6 is fitted similarly as for Set4 but without considering the $g^{(l)}_{n}$ term. RMS deviations between the observed and calculated magnetic moments are also listed in Table \ref{gfactor} with the free $g$ factors and the six sets of effective $g$ factors for comparison. Comparing with the results with free $g$ factors, all results from the six sets of effective $g$ factors much reduce the RMS deviations, which indicate strong core polarization effects on the $g$ factors.

\begin{table*}
\caption{\label{mm} The calculated and observed $44$ magnetic moments used in fitting. Observed data are taken from Ref.~\cite{Stone2005}. L$_{p}$ (L$_{n}$), S$_{p}$ (S$_{n}$), and t$_{p}$ (t$_{n}$) are the proton (neutron) orbital angular momentum, spin, and tensor part of the nuclear matrix elements, respectively. Set1, Set2, Set3, Set4, and Set5 are shell model calculations with six sets of effective $g$ factors listed in Table \ref{gfactor}. The unit for magnetic moments is $\mu_{N}$.}
\begin{ruledtabular}
\begin{tabular}{cccc cccc cccc cccc c}
&nuclides & J$^{\pi}$ & $l_{p}$& $l_{n}$& $s_{p}$& $s_{n}$ & $t_{p}$& $t_{n}$&free & Set1 & Set2 & Set3 & Set4 & Set5 & Set6 & Expt.\\ \hline
$N\geq126$&$^{211}$Pb&9/2$^{+}$  & 0.00  &4.02  &0.00  &0.48   &0.00  &0.37  & -1.83 &-1.09 &-1.10 &-1.07 &-1.43 &-1.43 &-1.44 & 1.4037(8)     \\
       &$^{210}$Pb  &   6$^{+}$  & 0.00  &5.38  &0.00  &0.62   &0.00  &0.48  & -2.35 &-1.40 &-1.42 &-1.37 &-1.86 &-1.85 &-1.85 & 1.87(9)       \\
       &$^{210}$Pb  &   8$^{+}$  & 0.00  &7.17  &0.00  &0.83   &0.00  &0.65  & -3.18 &-1.88 &-1.91 &-1.85 &-2.50 &-2.49 &-2.49 & 2.50(6)       \\
       &$^{209}$Pb  & 9/2$^{+}$  & 0.00  &4.00  &0.00  &0.50   &0.00  &0.36  & -1.91 &-1.14 &-1.15 &-1.11 &-1.48 &-1.48 &-1.50 & 1.4735(16)    \\
       &$^{208}$Tl  &   5$^{+}$  & 0.05  &3.98  &0.47  &0.50   &-0.25 &0.36  & 0.76  &0.67  &0.54  &0.45  &0.31  &0.29  &0.31  & +0.292(13)    \\
       &$^{207}$Tl  & 1/2$^{+}$  & 0.00  &0.00  &0.50  &0.00   &0.00  &0.00  & 2.79  &1.87  &1.87  &1.78  &1.86  &1.86  &1.86  & +1.876(5)     \\
       &$^{206}$Hg  &   5$^{-}$  & 4.89  &0.00  &0.11  &0.00   &0.03  &0.00  & 5.51  &5.47  &5.49  &5.41  &5.45  &5.45  &5.45  & +5.45(5)      \\ \hline
$N<126$&$^{207}$Pb  & 1/2$^{-}$  & 0.00  &0.67  &0.00  &-0.17  &0.00  &-0.67 & 0.64  &0.41  &0.44  &0.54  &0.35  &0.41  &0.33  & +0.592583(9)  \\
       &$^{207}$Pb  & 5/2$^{-}$  & 0.00  &2.86  &0.00  &-0.36  &0.00  &-0.57 & 1.37  &0.90  &0.92  &0.92  &0.76  &0.82  &0.71  & +0.80(3)      \\
       &$^{206}$Pb  &   7$^{-}$  & 0.00  &6.79  &0.00  &0.21   &0.00  &-0.19 & -0.79 &-0.39 &-0.38 &-0.33 &-0.38 &-0.37 &-0.41 & -0.152(3)     \\
       &$^{206}$Pb  &  12$^{+}$  & 0.00  &11.08 &0.00  &0.92   &0.00  &0.74  & -3.53 &-2.05 &-2.08 &-2.05 &-1.85 &-1.92 &-1.84 & -1.80(2)      \\
       &$^{205}$Pb  & 5/2$^{-}$  & 0.00  &2.83  &0.00  &-0.33  &0.00  &-0.58 & 1.28  &0.84  &0.87  &0.87  &0.71  &0.77  &0.67  & +0.7117(4)    \\
       &$^{205}$Pb  &13/2$^{+}$  & 0.00  &6.02  &0.00  &0.48   &0.00  &0.41  & -1.84 &-1.06 &-1.08 &-1.07 &-0.96 &-1.00 &-0.96 & -0.98(4)     \\
       &$^{205}$Pb  &25/2$^{-}$  & 0.00  &11.92 &0.00  &0.58   &0.00  &0.18  & -2.20 &-1.20 &-1.21 &-1.17 &-1.12 &-1.14 &-1.15 & -0.845(14)   \\
       &$^{205}$Pb  &33/2$^{+}$  & 0.00  &15.23 &0.00  &1.27   &0.00  &1.02  & -4.86 &-2.82 &-2.86 &-2.81 &-2.54 &-2.64 &-2.53 & -2.44(8)     \\
       &$^{203}$Pb  & 5/2$^{-}$  & 0.00  &2.82  &0.00  &-0.32  &0.00  &-0.58 & 1.21  &0.80  &0.82  &0.83  &0.68  &0.73  &0.63  & +0.6864(5)   \\
       &$^{203}$Pb  &25/2$^{-}$  & 0.00  &12.10 &0.00  &0.40   &0.00  &0.12  & -1.52 &-0.77 &-0.78 &-0.79 &-0.75 &-0.76 &-0.79 & -0.74(4)     \\
       &$^{202}$Pb  &   9$^{-}$  & 0.00  &8.89  &0.00  &0.11   &0.00  &-0.14 & -0.41 &-0.12 &-0.12 &-0.13 &-0.16 &-0.15 &-0.21 & -0.2276(7)   \\
       &$^{202}$Pb  &  16$^{+}$  & 0.00  &15.68 &0.00  &0.32   &0.00  &0.04  & -1.23 &-0.53 &-0.54 &-0.58 &-0.56 &-0.57 &-0.64 & -0.67(16)    \\
       &$^{202}$Pb  &  19$^{-}$  & 0.00  &18.17 &0.00  &0.83   &0.00  &0.50  & -3.17 &-1.71 &-1.73 &-1.74 &-1.60 &-1.65 &-1.65 & -1.88(6)     \\
       &$^{201}$Pb  & 5/2$^{-}$  & 0.00  &2.80  &0.00  &-0.30  &0.00  &-0.58 & 1.15  &0.76  &0.79  &0.80  &0.64  &0.70  &0.60  & +0.6753(5)   \\
       &$^{201}$Pb  &25/2$^{-}$  & 0.00  &12.10 &0.00  &0.40   &0.00  &0.11  & -1.53 &-0.78 &-0.78 &-0.78 &-0.75 &-0.76 &-0.80 & -0.79(4)     \\
       &$^{201}$Pb  &29/2$^{-}$  & 0.00  &14.00 &0.00  &0.50   &0.00  &0.21  & -1.90 &-0.98 &-0.99 &-1.00 &-0.94 &-0.96 &-0.99 & -1.011(6)    \\
       &$^{200}$Pb  &   7$^{-}$  & 0.00  &6.76  &0.00  &0.24   &0.00  &0.01  & -0.91 &-0.47 &-0.47 &-0.45 &-0.45 &-0.45 &-0.47 & -0.21(10)    \\
       &$^{200}$Pb  &   9$^{-}$  & 0.00  &8.88  &0.00  &0.12   &0.00  &-0.14 & -0.45 &-0.15 &-0.14 &-0.15 &-0.18 &-0.17 &-0.23 & -0.258(9)    \\
       &$^{200}$Pb  &  12$^{+}$  & 0.00  &11.14 &0.00  &0.86   &0.00  &0.76  & -3.31 &-1.91 &-1.94 &-1.93 &-1.73 &-1.80 &-1.72 & -1.849(12)   \\
       &$^{205}$Tl  & 1/2$^{+}$  & -0.08 &0.11  &0.47  &0.00   &0.01  &-0.01 & 2.51  &1.65  &1.66  &1.58  &1.68  &1.69  &1.68  & +1.63821461(12)\\
       &$^{205}$Tl  & 5/2$^{+}$  & 0.19  &1.87  &0.45  &-0.01  &0.19  &-0.07 & 2.73  &1.92  &2.01  &1.96  &1.93  &1.98  &1.91  & +2.0(3)        \\
       &$^{205}$Tl  &25/2$^{+}$  & 4.96  &6.89  &0.50  &0.15   &0.38  &-0.13 & 7.15  &6.72  &6.90  &6.84  &6.70  &6.80  &6.69  & +6.80(10)      \\
       &$^{204}$Tl  &   2$^{-}$  & 0.19  &2.20  &-0.20 &-0.19  &-0.21 &-0.31 & -0.16 &-0.04 &-0.12 &-0.14 &-0.13 &-0.15 &-0.16 & 0.09(1)        \\
       &$^{203}$Tl  & 1/2$^{+}$  & -0.05 &0.07  &0.47  &0.01   &0.01  &-0.01 & 2.57  &1.70  &1.70  &1.62  &1.73  &1.73  &1.73  & +1.62225787(12)\\
       &$^{203}$Tl  & 3/2$^{+}$  & 1.27  &0.53  &-0.29 &-0.01  &-0.53 &-0.02 & -0.28 &0.28  &0.05  &-0.03 &0.25  &0.12  &0.25  & 0.0(2)         \\
       &$^{202}$Tl  &   2$^{-}$  & 0.13  &2.14  &-0.14 &-0.13  &-0.14 &-0.24 & -0.16 &-0.05 &-0.10 &-0.11 &-0.12 &-0.13 &-0.14 & 0.06(1)        \\
       &$^{202}$Tl  &   7$^{+}$  & 0.05  &6.02  &0.45  &0.48   &0.26  &0.39  & 0.74  &0.68  &0.78  &0.76  &0.81  &0.83  &0.81  & +0.90(4)       \\
       &$^{201}$Tl  & 1/2$^{+}$  & -0.04 &0.06  &0.47  &0.01   &0.01  &-0.01 & 2.59  &1.72  &1.72  &1.64  &1.75  &1.75  &1.75  & +1.605(2)      \\
       &$^{200}$Tl  &   2$^{-}$  & 0.05  &1.99  &-0.02 &-0.02  &-0.05 &-0.10 & -0.01 &0.04  &0.02  &0.02  &0.02  &0.02  &0.00  & 0.04(1)        \\
       &$^{205}$Hg  & 1/2$^{-}$  & -0.03 &0.69  &0.00  &-0.16  &-0.01 &-0.62 & 0.61  &0.38  &0.40  &0.50  &0.32  &0.38  &0.30  & +0.60089(10)   \\
       &$^{204}$Hg  &   2$^{+}$  & 0.73  &1.26  &0.01  &0.00   &-0.05 &-0.04 & 0.80  &0.82  &0.80  &0.78  &0.81  &0.80  &0.80  & +0.9(2)        \\
       &$^{203}$Hg  & 5/2$^{-}$  & 0.19  &2.63  &0.00  &-0.32  &-0.01 &-0.54 & 1.43  &1.01  &1.03  &1.02  &0.89  &0.94  &0.84  & +0.84895(13)   \\
       &$^{202}$Hg  &   2$^{+}$  & 0.76  &1.23  &0.01  &0.00   &-0.04 &-0.03 & 0.80  &0.83  &0.81  &0.79  &0.82  &0.81  &0.81  & +0.78(6)       \\
       &$^{202}$Hg  &   4$^{+}$  & 1.12  &2.85  &0.02  &0.01   &-0.06 &-0.07 & 1.18  &1.24  &1.22  &1.18  &1.21  &1.20  &1.20  & 1.4(3)         \\
       &$^{201}$Hg  & 3/2$^{-}$  & 0.03  &1.07  &0.00  &0.04   &0.00  &0.12  &-1.50  &-0.91 &-0.92 &-0.82 &-0.80 &-0.81 &-0.77 &-0.5602257(14)  \\
       &$^{200}$Hg  &   2$^{+}$  & 0.73  &1.25  &0.01  &0.01   &-0.04 &-0.03 & 0.74  &0.78  &0.76  &0.75  &0.77  &0.76  &0.76  & +0.65(5)       \\
       &$^{200}$Hg  &   4$^{+}$  & 0.97  &2.99  &0.01  &0.02   &-0.05 &-0.07 & 0.96  &1.04  &1.03  &1.00  &1.02  &1.01  &1.01  & 1.02(17)       \\
       &$^{200}$Au  &  12$^{-}$  & 4.98  &6.06  &0.48  &0.48   &0.36  &0.40  & 5.83  &5.89  &6.04  &5.96  &5.98  &6.03  &6.00  &  5.90(9)       \\
RMS    &            &            &       &      &      &       &      &      & 0.77  & 0.21  & 0.20  &0.19  & 0.13  & 0.12  &0.13  &              \\
\end{tabular}
\end{ruledtabular}
\end{table*}

Any set of the effective $g$ factors shown in Table \ref{mm} successfully reproduces the experimentally observed $44$ magnetic moments. It is reasonable to use them for further investigations on other magnetic moments and M1 transitions. The magnetic moments are generally sensitive to the single-particle configurations. The nice agreements between observed and calculated magnetic moments show that the present Hamiltonian reasonably accounts for the dominant configurations of considered states in Table \ref{mm}.

As seen in Table \ref{gfactor} and Table \ref{mm}, total RMS deviation and deviations between observed and calculated magnetic moments in $N\geq126$ nuclei are much reduced if separate sets of effective $g$ factors are used for magnetic moments with $N\geq126$ and $N<126$. The calculated magnetic moments with the effective $g$ factors obtained from global fitting, Set1, Set2, and Set3, have certain deviations from the observations for the neutron-rich nuclei with $N\geq126$, such as $^{209,210,211}$Pb and $^{205}$Tl shown worthwhile in Table \ref{mm}. If the effective $g$ factors are separately fitted for $N\geq126$ and $N<126$ nuclei, the results from Set4, Set5, and Set6 almost exactly reproduce the seven observed magnetic moments of $^{209,210,211}$Pb, $^{205,207}$Tl, and $^{206}$Hg. Figure \ref{mmfig} presents a comparison between calculated results and observed data. The contribution of the $g^{(t)}$ term is so small that the Set1 results are similar to those of Set3 with the $g^{(t)}$ terms. After the separate consideration of $N\geq126$ and $N<126$ nuclei, the results with Set5 give a better agreement with the data compared with those of Set2. It is worthwhile to discuss the values of effective $g$ factors to understand the origin of the difference.

In general, free $g^{(s)}$ factors should be quenched to reproduce the observed magnetic moments because of the core polarization effect. For the present results, quenching with factors of $64-68\%$ and $52-63\%$ (except that of set 6) are found for $g^{(s)}_{p}$ and $g^{(s)}_{n}$, respectively, which show stronger quenching comparing with those around $70\%$ in the $f_{5/2}pg_{9/2}$-shell region \cite{Honma2009}, and around $90\%$ in the $sd$-shell region \cite{Richter2008PRC} and $psd$-shell region \cite{yuan2012}. It is reasonable that a larger core corresponds to a stronger core polarization effect and quenching.

All six sets of the effective $g$ factors show a positive $\delta g^{(l)}_{p}$, defined as $g^{(l)}_{p,eff}=g^{(l)}_{p,free}+\delta g^{(l)}_{p}$, which is reasonable due to the meson exchange processes \cite{arima1986,towner1987}. A negative $\delta g^{(l)}_{n}$, defined as $g^{(l)}_{n,eff}=g^{(l)}_{n,free}+\delta g^{(l)}_{n}$, is found with large uncertainty in the set of the effective $g$ factors fitted to the magnetic moments with $N\geq126$. The other sets of the effective $g$ factors give a positive but tiny $\delta g^{(l)}_{n}$ except for Set6, which uses free $g^{(l)}_{n}$. For the magnetic moments with $N<126$, a negative $\delta g^{(l)}_{n}$ increases the RMS deviation dramatically. If $\delta g^{(l)}_{n}=-0.05$ is used for magnetic moments with $N<126$ in Set4, the total RMS deviation increases from 0.13 to 0.40.

Even if we use a free $g^{(l)}_{n}$ ($g^{(l)}_{n}=0$), the present data is successfully reproduced and the RMS deviation is reasonably small. If $g^{(l)}_{n}=0$ is used in Set4 and other effective $g$ factors are refitted, as shown for Set6 in Table \ref{gfactor}, the RMS deviation increases little from 0.12 to 0.13. In addition, all other effective $g$ factors change little except for the $g^{(s)}_{n}$ factor for $N\geq126$ nuclei: $g^{(s)}_{n}$ changes from -2.386(0.603) $\mu_{n}$ to -3.001(0.020) $\mu_{n}$, which is much different from that used for $N<126$ nuclei. The different effective $g^{(s)}_{n}$ term for $N\geq126$ and $N<126$ nuclei is the main reason that the RMS deviation is much reduced when the two situations are separately considered.

The contribution of the tensor term to the magnetic moments is rarely discussed. In the $sd$-shell region, the inclusion of the tensor terms changes little the total RMS deviation between observed and calculated magnetic moments and M1 transitions \cite{Richter2008PRC}. The present work shows similar results that the inclusion of the tensor terms change the RMS deviation little, as seen from Table \ref{gfactor}. The RMS deviations are close to each other among Set1, Set2, and Set3, and between Set4 and Set5.

After separately considering the $N\geq126$ and $N<126$ nuclei, results are presented with free fitting of the $g^{(l)}$ and $g^{(s)}$ terms (Set4), with fitting of the $g^{(t)}$ terms when the $g^{(l)}$ and $g^{(s)}$ term are fixed as Set4 (Set5), and with fitting of the $g^{(l)}$ and $g^{(s)}$ terms when the $g^{(l)}_{n}$ term is constrained to be zero (Set6). The results of free fitting of all the $g^{(l)}$, $g^{(s)}$, and $g^{(t)}$ terms are not presented, because there are only seven nuclei with $N\geq126$, too few to fit six effective $g$ factors. For $N<126$ nuclei, the free fitting of all the $g^{(l)}$, $g^{(s)}$, and $g^{(t)}$ terms results in $g^{(s)}_{n}=-1.116(0.168)$ and $g^{(t)}_{n}=-0.761(0.122)$, which are quite different from those in the six sets and largely quenched from the free values, while the other four factors are not much different from those in the six sets. It seems the $g^{(t)}_{n}$ term substitutes partial effect of $g^{(s)}_{n}$ term. Similar but less pronounced effects are found in Set3 and in $sd$-shell region in Ref. \cite{Richter2008PRC}. It demands further and systematic investigations on the relationship between the $g^{(s)}_{n}$ and $g^{(t)}_{n}$ terms.

For the $3/2^{-}$ state in $^{201}$Hg, unlike its quadrupole moment, the first calculated $3/2^{-}$ state gives the closest value of magnetic moment to the observed one, while the second and third calculated $3/2^{-}$ states have both positive magnetic moments. Considering that the quadrupole moment of the third calculated $3/2^{-}$ state mostly agrees with the experimental data, it is difficult to identify which calculated state corresponds to the observed ground state. Actually, configuration mixing is quite strong for the three calculated $3/2^{-}$ states, where the percentage of the largest component in these three states is less than $6\%$.

\begin{table*}
\caption{\label{mm1} The same as Table \ref{mm}, but for magnetic moments with observed data not used in fitting. The results of both the first and second calculated states are presented for the corresponding states in $^{202-204}$Pb and $^{205}$Tl. }
\begin{ruledtabular}
\begin{tabular}{ccccccccccccccc}
nuclides & J$^{\pi}$ & $l_{p}$& $l_{n}$& $s_{p}$& $s_{n}$ & $t_{p}$& $t_{n}$ & Set1 & Set2 & Set3 & Set4 & Set5 & Set6 & Expt.\\ \hline
$^{206}$Pb  &     2$^{+}$   &0.00  &2.05  &0.00  &-0.05 &0.00  &-0.07 &0.15  &0.16  &0.14  &0.12  &0.13  &0.10  &  $<$0.03      \\
$^{204}$Pb  &     2$^{+}$   &0.00  &2.04  &0.00  &-0.04 &0.00  &-0.12 &0.12  &0.12  &0.12  &0.09  &0.10  &0.07  &  $<$0.02      \\
$^{206}$Tl  &   (5)$^{+}$   &4.38  &-0.18 &0.48  &0.33  &0.37  &0.43  &5.54  &5.69  &5.62  &5.64  &5.68  &5.68  & +4.27(6)      \\
$^{204}$Tl  &   (7)$^{+}$   &0.07  &6.01  &0.45  &0.48  &0.27  &0.39  &0.68  &0.79  &0.78  &0.81  &0.84  &0.81  & +1.187(6)     \\
$^{203}$Tl  &   5/2$^{+}$   &0.27  &1.76  &0.46  &0.01  &0.19  &-0.07 &2.00  &2.09  &2.04  &2.01  &2.07  &2.00  & +2.6(11)      \\
$^{204}$Pb  &     4$^{+}$   &0.00  &4.34  &0.00  &-0.34 &0.00  &-0.58 &0.87  &0.90  &0.88  &0.73  &0.78  &0.67  &  +0.225(4)    \\
            &               &0.00  &4.11  &0.00  &-0.11 &0.00  &-0.64 &0.34  &0.36  &0.45  &0.27  &0.33  &0.23  &  +0.225(4)    \\
$^{203}$Pb  &  21/2$^{+}$   &0.00  &10.45 &0.00  &0.05  &0.00  &-0.15 &0.03  &0.04  &-0.01 &-0.04 &-0.03 &-0.11 &  -0.64(2)     \\
            &               &0.00  &10.17 &0.00  &0.33  &0.00  &-0.18 &-0.63 &-0.62 &-0.56 &-0.61 &-0.59 &-0.65 &  -0.64(2)     \\
$^{202}$Pb  &     4$^{+}$   &0.00  &4.21  &0.00  &-0.21 &0.00  &-0.42 &0.58  &0.59  &0.58  &0.47  &0.51  &0.42  &  +0.008(16)   \\
            &               &0.00  &4.13  &0.00  &-0.13 &0.00  &-0.81 &0.38  &0.42  &0.54  &0.31  &0.38  &0.27  &  +0.008(16)   \\
$^{205}$Tl  &   3/2$^{+}$   &1.20  &0.60  &-0.28 &-0.02 &-0.53 &-0.02 &0.24  &0.00  &-0.07 &0.20  &0.08  &0.20  & -0.8(5)       \\
            &               &0.44  &1.15  &-0.14 &0.05  &-0.37 &0.03  &-0.15 &-0.32 &-0.37 &-0.16 &-0.25 &-0.16 & -0.8(5)       \\
\end{tabular}
\end{ruledtabular}
\end{table*}

Some other experimental data on magnetic moments are available in this region in nuclei with $A\geq200$. They are not discussed in Table \ref{mm} but listed in Table \ref{mm1}. Some of these magnetic moments have no exact experimental values, such as the $2^{+}_{1}$ states of $^{204,206}$Pb, or uncertainties on spins, such as the $(5)^{+}_{1}$ state of $^{206}$Tl and the $(7)^{+}_{1}$ state of $^{204}$Tl, or large uncertainty in observed value, such as $5/2^{+}_{1}$ state of $^{203}$Tl, or large deviation between observed and calculated values, such as the $4^{+}_{1}$ states of $^{202,204}$Pb, the $23/2^{+}_{1}$ state of $^{203}$Pb, and the $3/2^{+}_{1}$ state of $^{205}$Tl.

For magnetic moments of the $2^{+}_{1}$ states of $^{204,206}$Pb, the $(5)^{+}_{1}$ state of $^{206}$Tl, the $(7)^{+}_{1}$ state of $^{204}$Tl, and the $5/2^{+}_{1}$ state of $^{203}$Tl, the present calculations provide reasonable results, which are not further discussed because of the lack of experimental information. As shown in Table \ref{mm1}, the calculated magnetic moments of the second corresponding states with the same spins and parities provide more reasonable descriptions of those of the $4^{+}_{1}$ states of $^{202,204}$Pb, the $23/2^{+}_{1}$ state of $^{203}$Pb, and the $3/2^{+}_{1}$ state of $^{205}$Tl. The corresponding yrare states locate not much higher than these yrast states. It is possible that the calculations do not reproduce the order of the yrast and yrare states because of the theoretical uncertainties.

The magnetic moment of the $3^{-}_{1}$ state in $^{208}$Pb is +1.9(2) $\mu_{N}$~\cite{Stone2005}, which is far from the calculated value, which is around -0.8 $\mu_{N}$. Such deviation indicates that the present model space is not large enough for this $3^{-}$ state, which is considered to be a surface vibration state with octupole type \cite{lane1960,carter1960} and needs to be described in a much larger model space. In general, it is difficult for the shell model to describe such octupole states, as discussed in Sec. \ref{sec:level5}.

\begin{table*}
\caption{\label{mm2} The same as Table \ref{mm}, but for predicted magnetic moments without observed data.}
\begin{ruledtabular}
\begin{tabular}{cccccccccccccc}
nuclides &J$^{\pi}$ & $l_{p}$& $l_{n}$& $s_{p}$& $s_{n}$ & $t_{p}$& $t_{n}$ & Set1 & Set2 & Set3 & Set4 & Set5& Set6   \\ \hline
$^{217}$Pb  &   $9/2^+$  &0.00  &4.08 &0.00 &0.42 &0.00  &0.38 &-0.96 &-0.97 &-0.96 &-1.31 &-1.30 &  -1.27 \\
$^{215}$Pb  &   $9/2^+$  &0.00  &4.06 &0.00 &0.44 &0.00  &0.38 &-1.00 &-1.01 &-0.99 &-1.34 &-1.34 &  -1.32 \\
$^{213}$Pb  &   $9/2^+$  &0.00  &4.04 &0.00 &0.46 &0.00  &0.37 &-1.04 &-1.06 &-1.03 &-1.39 &-1.38 &  -1.38 \\
$^{216}$Pb  &   $  8^+$  &0.00  &7.26 &0.00 &0.74 &0.00  &0.68 &-1.66 &-1.69 &-1.67 &-2.29 &-2.28 &  -2.22 \\
$^{214}$Pb  &   $  8^+$  &0.00  &7.23 &0.00 &0.77 &0.00  &0.67 &-1.74 &-1.77 &-1.73 &-2.36 &-2.35 &  -2.31 \\
$^{212}$Pb  &   $  8^+$  &0.00  &7.20 &0.00 &0.80 &0.00  &0.66 &-1.81 &-1.84 &-1.79 &-2.43 &-2.43 &  -2.40 \\
$^{213}$Tl  &   $1/2^+$  &-0.03 &0.04 &0.49 &0.00 &0.00  &0.01 &1.81  &1.81  & 1.72 & 1.79 &1.79  &  1.81  \\
$^{211}$Tl  &   $1/2^+$  &-0.03 &0.04 &0.49 &0.00 &0.00  &0.01 &1.81  &1.81  & 1.72 & 1.79 &1.79  &  1.80  \\
$^{209}$Tl  &   $1/2^+$  &-0.02 &0.03 &0.49 &0.00 &0.00  &0.00 &1.82  &1.82  & 1.73 & 1.81 &1.81  &  1.82  \\
$^{212}$Tl  &   $  5^+$  &-0.01 &4.07 &0.49 &0.45 &-0.03 &0.38 &0.79  &0.76  & 0.69 & 0.43 &0.43  &  0.46  \\
$^{210}$Tl  &   $  5^+$  &0.01  &4.03 &0.49 &0.48 &-0.13 &0.37 &0.74  &0.66  & 0.59 & 0.38 &0.37  &  0.39  \\
$^{209}$Hg  &   $9/2^+$  &0.04  &3.99 &0.00 &0.47 &0.01  &0.36 &-1.03 &-1.04 &-1.00 &-1.37 &-1.37 &  -1.38 \\
$^{207}$Hg  &   $9/2^+$  &0.04  &3.96 &0.00 &0.50 &0.01  &0.36 &-1.09 &-1.10 &-1.05 &-1.43 &-1.43 &  -1.45 \\
\end{tabular}
\end{ruledtabular}
\end{table*}

Based on the six sets of the effective $g$ factors, magnetic moments of some neutron-rich Pb, Tl, and Hg isotopes are predicted in Table \ref{mm2}. Because the nuclei in Table \ref{mm2} are all with $N>126$, magnetic moments predicted with Set4, Set5, and Set6 are preferentially recommended. The values of magnetic moments of the ground states, $9/2^{+}$, of $^{211,213,215,217}$Pb and $^{207,209}$Hg are quite similar to each other, which are dominated by a single $1g_{9/2}$ neutron configuration. The $\nu(1g_{9/2})^2$, $\pi(2s_{1/2})^{-1}$, and $\pi(2s_{1/2})^{-1}\nu(1g_{9/2})^1$ configurations contribute dominantly to the magnetic moments of the $8^{+}_1$ excited states of $^{210,212,214,216}$Pb, the $1/2^{+}$ ground states of $^{207,209,211,213}$Tl, and the $5^{+}$ ground states of $^{208,210,212}$Tl, respectively.

\section{\label{sec:level8}Summary}

In summary, systematic shell-model investigations are performed for the south region of $^{208}$Pb. The model space includes five proton orbits and thirteen neutron orbits. The two-body matrix elements (TBME) are partly taken from the existing KHHE and KHPE Hamiltonians and partly calculated through the monopole based universal interaction V$_{\text{MU}}$ plus spin-orbit interaction. The newly constructed Hamiltonian can well reproduce binding energies, levels (including those of neutron core-excited states and isomeric states), electric and magnetic properties of nuclei in the south proximity of $^{208}$Pb (such as Pb, Tl, Hg, Au, Pt, and Ir isotopes around the $N = 126$ shell).

Based on the reasonable description of known properties, the present Hamiltonian predicts some unknown binding energies in nuclei around $N = 126$ and possible isomeric states in neutron-rich Pb, Tl, and Hg isotopes. The $N = 126$ shell closure is predicted to be kept unchanged from $Z = 82$ to $68$ with minor reduction of the gap.

We adopted the effective charges $e_{p}=1.80$ and $e_{n}=0.80$ for this mass region through the systematic investigations on $32$ electric quadrupole properties. For magnetic dipole properties, it is recommended to use different sets of the effective $g$ factors to calculate the magnetic moments of the $N\geq126$ and $N<126$ nuclei, while the reason is also clarified. If the tensor terms of the M1 operator are not taken into account as many investigations, effective $g$ factors, Set4 and Set6 in Table \ref{gfactor} (each includes two sets for the $N\geq126$ and $N<126$ nuclei, respectively), show reasonable agreement with the experimental values through the present studies on $44$ magnetic moments.

Recently, many $\beta$-decay properties were measured in neutron-rich Au, Hg, Tl, Pb, and Bi isotopes, including decay half-lives, $\beta$ delayed $\gamma$ transitions, and neutron emissions~\cite{Morales2014,Caballero2016,Caballero2017}. It is expected that the present Hamiltonian can be used to investigate these $\beta$-decay properties, especially for nuclei around $N=126$ shell closure. As discussed in previous works, the first-forbidden transitions in $\beta$ decay are significant in the south of $^{208}$Pb~\cite{Suzuki2012,Nishimura2016}. A systematic study based on the present Hamiltonian would be helpful for investigations on the $r$ process, including both Gamow-Teller and first-forbidden transitions.

\section{\label{sec:level9}Acknowledgement}

This work has been supported by the Guangdong Major Project of Basic and Applied Basic Research under Grant No. 2021B0301030006, the National Natural Science Foundation of China under Grant Nos.~11775316, 11961141004, 11675225, 11635003, and 11735017, the National Key Research and Development Program of China under Grant Nos. 2018YFA0404402 and 2018YFB1900405, the Tip-top Scientific and Technical Innovative Youth Talents of Guangdong special support program under Grant No.~2016TQ03N575, the computational resources from SYSU and National Supercomputer Center in Guangzhou, and STFC(UK). NS and TO acknowledges acknowledge the support from MEXT as "Priority Issue on post-K computer" (Elucidation of the Fundamental Laws and Evolution of the Universe).

\section{\label{sec:level10}Appendix}

The calculated binding energies are listed in the appendix for Hg, Au, Pt, Ir, Os, Re, and W isotopes based on three Hamiltonians, as shown in Tables \ref{BE2-1} and \ref{BE2-2}. Besides the present results BE$_{present}$, results from two Hamiltonians, BE$_{KHHE}$ and BE$_{Ste11}$, with the same proton-neutron and neutron-neutron parts are also presented to show the effect of the modification of the proton-proton interaction. BE$_{KHHE}$ and BE$_{Ste11}$ are calculated with the proton-proton interaction from the original KHHE and the modified one mentioned in Ref.~\cite{Steer2011} (Ste11 means Steer \emph{et al.}, 2011), respectively. The modifications in Ref.~\cite{Steer2011} aimed to give better descriptions on the proton-hole states of $^{205}$Au and $^{203}$Ir. As described in Sec. \ref{sec:level2}, BE$_{present}$ are calculated with the proton-proton interaction from KHHE with $0.1$ MeV added to interactions of $1d_{3/2}$, $1d_{5/2}$, $0g_{7/2}$, and $0h_{11/2}$ orbits, respectively. Because only the proton-proton interaction is different among these three Hamiltonians and the single-particle energies of $^{207}$Tl are fixed to the observed data, all these Hamiltonians give the same results for Pb and Tl isotopes, which are already presented in Table \ref{BE1-1}.

BE$_{present}$ are in good agreement with the $23$ observed binding energies of Hg, Au, and Pt isotopes taken from AME2020~\cite{AME2020}. If the proton-proton interaction is not enlarged, binding energies of $^{198-203}$Au, $^{198-202}$Pt, $^{199}$Ir and are overbound as in the BE$_{KHHE}$ and BE$_{Ste11}$ results compared with the observed data. It is also found that without the present modification, BE$_{KHHE}$ and BE$_{Ste11}$ generally present overbound results compared with those predicted by AME2020, especially when proton number becomes smaller. Because BE$_{KHHE}$ and BE$_{Ste11}$ overestimate the observed binding energies of Au and Pt isotopes, it is reasonable to assume that they also overestimate the unmeasured ones. BE$_{present}$ give more reasonable predictions compared with those in AME2020.

\begin{table*}
\caption{\label{BE2-1} The comparison between the calculated and observed binding energies (unit in MeV). Experimental data are taken from AME2020~\cite{AME2020}.}
\begin{ruledtabular}
\begin{tabular}{cccccccccccc}
nuclide&BE$_{KHHE}$&BE$_{Ste11}$&BE$_{present}$&BE$_{Expt.}$&$\Delta$BE$_{KHHE}$&$\Delta$BE$_{Ste11}$&$\Delta$BE$_{present}$ \\ \hline
$^{208}$Hg & 1629.556 & 1629.552 & 1629.480 & 1629.512 & -0.044  & -0.040  & 0.032  \\
$^{207}$Hg & 1624.690 & 1624.687 & 1624.618 & 1624.662 & -0.028  & -0.025  & 0.044  \\
$^{206}$Hg & 1621.071 & 1621.071 & 1620.993 & 1621.049 & -0.022  & -0.022  & 0.056  \\
$^{205}$Hg & 1614.336 & 1614.330 & 1614.239 & 1614.320 & -0.016  & -0.010  & 0.081  \\
$^{204}$Hg & 1608.644 & 1608.627 & 1608.548 & 1608.651 & 0.007   & 0.024   & 0.103  \\
$^{203}$Hg & 1601.164 & 1601.144 & 1601.067 & 1601.159 & -0.005  & 0.015   & 0.092  \\
$^{202}$Hg & 1595.241 & 1595.219 & 1595.141 & 1595.164 & -0.077  & -0.055  & 0.023  \\
$^{201}$Hg & 1587.449 & 1587.423 & 1587.349 & 1587.410 & -0.039  & -0.013  & 0.061  \\
$^{200}$Hg & 1581.297 & 1581.271 & 1581.194 & 1581.179 & -0.118  & -0.092  & -0.015  \\
$^{199}$Hg & 1573.246 & 1573.213 & 1573.149 & 1573.151 & -0.095  & -0.062  & 0.002  \\
$^{198}$Hg & 1566.796 & 1566.769 & 1566.690 & 1566.487 & -0.309  & -0.282  & -0.203  \\
$^{203}$Au & 1600.051 & 1600.020 & 1599.855 & 1599.816 & -0.235  & -0.204  & -0.039 \\
$^{202}$Au & 1593.147 & 1593.097 & 1592.925 & 1592.954 & -0.193  & -0.143  & 0.029  \\
$^{201}$Au & 1587.296 & 1587.248 & 1587.087 & 1586.930 & -0.366  & -0.318  & -0.157 \\
$^{200}$Au & 1580.001 & 1579.943 & 1579.783 & 1579.698 & -0.303  & -0.245  & -0.085 \\
$^{199}$Au & 1573.946 & 1573.889 & 1573.729 & 1573.481 & -0.465  & -0.408  & -0.248 \\
$^{198}$Au & 1566.332 & 1566.265 & 1566.108 & 1565.896 & -0.436  & -0.369  & -0.212 \\
$^{202}$Pt & 1592.517 & 1592.390 & 1591.962 & 1592.075 & -0.442  & -0.315  & 0.113  \\
$^{201}$Pt & 1585.684 & 1585.547 & 1585.121 & 1585.052 & -0.632  & -0.495  & -0.069  \\
$^{200}$Pt & 1580.513 & 1580.364 & 1579.937 & 1579.840 & -0.673  & -0.524  & -0.097 \\
$^{199}$Pt & 1573.293 & 1573.135 & 1572.711 & 1572.558 & -0.735  & -0.577  & -0.153 \\
$^{198}$Pt & 1567.860 & 1567.701 & 1567.271 & 1567.002 & -0.858  & -0.699  & -0.269 \\
$^{199}$Ir & 1571.507 & 1571.294 & 1570.561 & 1570.351 & -1.156  & -0.943  & -0.210 \\
RMS        &          &          &          &          &  0.443  &  0.358  &  0.129 \\
\end{tabular}
\end{ruledtabular}
\end{table*}

\begin{table*}
\caption{\label{BE2-2} The comparison between the calculated binding energies and binding energies predicted in AME2020~\cite{AME2020} (unit in MeV).}
\begin{ruledtabular}
\begin{tabular}{cccccccccccc}
nuclide&BE$_{KHHE}$&BE$_{Ste11}$&BE$_{present}$&BE$_{AME}$&$\Delta$BE$_{KHHE}$&$\Delta$BE$_{Ste11}$&$\Delta$BE$_{present}$ \\ \hline
$^{217}$Pb & 1675.141 & 1675.141 & 1675.141 & 1675.023 & -0.118  & -0.118  & -0.118   \\
$^{216}$Pb & 1671.749 & 1671.749 & 1671.749 & 1671.840 & 0.091   & 0.091   & 0.091    \\
$^{212}$Tl & 1649.379 & 1649.379 & 1649.379 & 1649.360 & -0.019  & -0.019  & -0.019   \\
$^{211}$Hg & 1641.215 & 1641.207 & 1641.145 & 1640.947 & -0.268 & -0.260  & 0.198    \\
$^{210}$Hg & 1637.875 & 1637.868 & 1637.801 & 1637.790 & -0.085  & -0.078  & 0.011    \\
$^{209}$Hg & 1633.051 & 1633.045 & 1632.980 & 1632.917 & -0.134  & -0.128  & -0.063   \\
$^{209}$Au & 1628.037 & 1628.016 & 1627.867 & 1627.274 & -0.763  & -0.742  & -0.593   \\
$^{208}$Au & 1623.455 & 1623.437 & 1623.296 & 1623.024 & -0.431  & -0.413  & -0.272   \\
$^{207}$Au & 1620.096 & 1620.083 & 1619.925 & 1619.568 & -0.528  & -0.515 & -0.357  \\
$^{206}$Au & 1615.499 & 1615.487 & 1615.346 & 1615.040 & -0.459  & -0.447  & -0.306   \\
$^{205}$Au & 1611.996 & 1611.993 & 1611.820 & 1611.300 & -0.696  & -0.693  & -0.520   \\
$^{204}$Au & 1605.657 & 1605.610 & 1605.441 & 1605.072 & -0.585  & -0.538  & -0.369   \\
$^{207}$Pt & 1614.735 & 1614.615 & 1614.225 & 1613.979 & -0.756  & -0.636  & -0.246   \\
$^{206}$Pt & 1611.613 & 1611.502 & 1611.098 & 1610.920 & -0.693  & -0.582  & -0.178    \\
$^{205}$Pt & 1607.072 & 1606.967 & 1606.562 & 1606.380 & -0.692  & -0.587  & -0.182    \\
$^{204}$Pt & 1603.812 & 1603.720 & 1603.291 & 1603.236 & -0.576  & -0.484  & -0.055    \\
$^{203}$Pt & 1597.645 & 1597.542 & 1597.105 & 1597.001 & -0.644  & -0.541  & -0.104    \\
$^{204}$Ir & 1597.002 & 1596.862 & 1596.142 & 1595.892 & -1.110  & -0.970  & -0.250   \\
$^{203}$Ir & 1593.890 & 1593.768 & 1593.038 & 1592.535 & -1.355  & -1.233  & -0.503   \\
$^{202}$Ir & 1588.082 & 1587.905 & 1587.182 & 1586.710 & -1.372  & -1.195  & -0.472   \\
$^{201}$Ir & 1583.037 & 1582.860 & 1582.132 & 1581.870 & -1.167  & -0.990  & -0.262   \\
$^{200}$Ir & 1576.617 & 1576.404 & 1575.699 & 1575.600 & -1.017  & -0.804  & -0.099   \\
$^{198}$Ir & 1564.841 & 1564.552 & 1563.834 & 1563.606 & -1.235  & -0.946  & -0.228   \\
$^{203}$Os & 1588.002 & 1587.683 & 1586.666 & 1586.242 & -1.760  & -1.441  & -0.424   \\
$^{202}$Os & 1585.095 & 1584.814 & 1583.752 & 1583.478 & -1.617  & -1.336  & -0.274    \\
$^{201}$Os & 1579.437 & 1579.125 & 1578.066 & 1577.649 & -1.788  & -1.476  & -0.417    \\
$^{200}$Os & 1574.854 & 1574.505 & 1573.440 & 1573.400 & -1.454  & -1.105  & -0.040    \\
$^{199}$Os & 1568.575 & 1568.180 & 1567.141 & 1566.926 & -1.649  & -1.254  & -0.215	\\
$^{202}$Re & 1577.062 & 1576.531 & 1575.202 &          &         &         &          \\
$^{201}$Re & 1574.248 & 1573.712 & 1572.396 &          &         &         &          \\
$^{200}$Re & 1569.016 & 1568.506 & 1567.081 &          &         &         &          \\
$^{199}$Re & 1564.531 & 1563.984 & 1562.566 & 1562.150 & -2.381  & -1.834  & -0.416   \\
$^{198}$Re & 1558.718 & 1558.151 & 1556.692 & 1556.478 & -2.240  & -1.673  & -0.214	\\
$^{201}$W  & 1567.529 & 1566.871 & 1564.982 &          &         &         &          \\
$^{200}$W  & 1564.967 & 1564.325 & 1562.421 &          &         &         &          \\
$^{199}$W  & 1559.803 & 1559.132 & 1557.215 &          &         &         &          \\
$^{198}$W  & 1555.738 & 1555.065 & 1553.076 &          &         &         &          \\
$^{197}$W  & 1549.963 & 1549.258 & 1547.252 & 1547.041 & -2.922  & -2.217  & -0.211   \\
RMS        &          &          &          &          &  1.222  &  0.984  &  0.293 \\
\end{tabular}
\end{ruledtabular}
\end{table*}


\begin{thebibliography}{99}
\bibitem{Mumpower2016} M.R. Mumpower, R. Surman, G.C. McLaughlin, and A. Aprahamian, Prog. Part. Nucl. Phys. {\bf 86}, 86 (2016).
\bibitem{Kajino2017} T. Kajino and G.J. Mathews, Rep. Prog. Phys. {\bf 80}, 084901 (2017).
\bibitem{Steer2011} S.J. Steer, Zs. Podoly\'{a}k, S. Pietri, M.G\'{o}rska, H. Grawe, K.H. Maier, P.H. Regan, D. Rudolph, A.B. Garnsworthy, R. Hoischen, J. Gerl, H.J. Wollersheim, F. Becker, P. Bednarczyk, L. C\'{a}ceres, P. Doornenbal, H. Geissel, J. Gr\c{e}bosz, A. Kelic, I. Kojouharov, N. Kurz, F. Montes, W. Prokopwicz, T. Saito, H. Schaffner, S. Tashenov, A. Heinz, M. Pf\"{u}tzner, T. Kurtukian-Nieto, G. Benzoni, A. Jungclaus, D.L. Balabanski, M. Bowry, C. Brandau, A. Brown, A.M. Bruce, W.N. Catford, I.J. Cullen, Zs. Dombr\'{a}di, M.E. Estevez, W. Gelletly, G. Ilie, J. Jolie, G.A. Jones, M. Kmiecik, F.G. Kondev, R. Kr\"{u}cken, S. Lalkovski, Z. Liu, A. Maj, S. Myalski, S. Schwertel, T. Shizuma, P. M. Walker, E. Werner-Malento, and O. Wieland, Phys Rev. C {\bf 84}, 044313 (2011).
\bibitem{Gottardo2012} A. Gottardo, J.J. Valiente-Dob\'{o}n, G. Benzoni, R. Nicolini, A. Gadea, S. Lunardi, P. Boutachkov, A.M. Bruce, M.G\'{o}rska, J. Grebosz, S. Pietri, Zs. Podoly\'{a}k, M. Pf\"{u}tzner, P.H. Regan, H. Weick, J. Alc\'{a}ntara N\'{u}\~{n}ez, A. Algora, N. Al-Dahan, G. de Angelis, Y. Ayyad, N. Alkhomashi, P.R.P. Allegro, D. Bazzacco, J. Benlliure, M. Bowry, A. Bracco, M. Bunce, F. Camera, E. Casarejos, M.L. Cortes, F.C.L. Crespi, A. Corsi, A.M. Denis Bacelar, A.Y. Deo, C. Domingo-Pardo, M. Doncel, Zs. Dombradi, T. Engert, K. Eppinger, G.F. Farrelly, F. Farinon, E. Farnea, H. Geissel, J. Gerl, N. Goel, E. Gregor, T. Habermann, R. Hoischen, R. Janik, S. Klupp, I. Kojouharov, N. Kurz, S.M. Lenzi, S. Leoni, S. Mandal, R. Menegazzo, D. Mengoni, B. Million, A.I. Morales, D.R. Napoli, F. Naqvi, C. Nociforo, A. Prochazka, W. Prokopowicz, F. Recchia, R.V. Ribas, M.W. Reed, D. Rudolph, E. Sahin, H. Schaffner, A. Sharma, B. Sitar, D. Siwal, K. Steiger, P. Strmen, T.P.D. Swan, I. Szarka, C.A. Ur, P.M. Walker, O. Wieland, H-J. Wollersheim, F. Nowacki, E. Maglione, and A.P. Zuker, Phys. Rev. Lett. {\bf 109}, 162502 (2012).
\bibitem{Gottardo2013} A. Gottardo, J.J. Valiente-Dob\'{o}n, G. Benzoni, A. Gadea, S. Lunardi, P. Boutachkov, A.M. Bruce, M. G\'{o}rska, J. Grebosz, S. Pietri, Zs. Podoly\'{a}k, M. Pf\"{u}tzner, P.H. Regan, H. Weick, J. Alc\'{a}ntara N\'{u}\~{n}ez, A. Algora, N. Al-Dahan, G. de Angelis, Y. Ayyad, N. Alkhomashi, P.R.P. Allegro, D. Bazzacco, J. Benlliure, M. Bowry, A. Bracco, M. Bunce, F. Camera, E. Casarejos, M.L. Cortes, F.C.L. Crespi, A. Corsi, A.M. Denis Bacelar, A.Y. Deo, C. Domingo-Pardo, M. Doncel, Zs. Dombradi, T. Engert, K. Eppinger, G.F. Farrelly, F. Farinon, E. Farnea, H. Geissel, J. Gerl, N. Goel, E. Gregor, T. Habermann, R. Hoischen, R. Janik, P.R. John, S. Klupp, I. Kojouharov, N. Kurz, S.M. Lenzi, S. Leoni, S. Mandal, R. Menegazzo, D. Mengoni, B. Million, V. Modamio, A.I. Morales, D.R. Napoli, F. Naqvi, R. Nicolini, C. Nociforo, A. Prochazka, W. Prokopowicz, F. Recchia, R.V. Ribas, M.W. Reed, D. Rudolph, E. Sahin, H. Schaffner, A. Sharma, B. Sitar, D. Siwal, K. Steiger, P. Strmen, T.P.D. Swan, I. Szarka, C.A. Ur, P.M. Walker, O. Wieland, H.-J. Wollersheim, Phys. Lett. B {\bf 725}, 292 (2013).

\bibitem{Miyatake2018} H. Miyatake, M. Wada, X.Y. Watanabe, Y. Hirayama, P. Schury, M. Ahmed, H. Ishiyama, S.C. Jeong, Y. Kakiguchi, S. Kimura, J.Y. Moon, M. Mukai, M. Oyaizu, and J.H. Park, AIP Conf. Proc. {\bf 1947}, 020018 (2018).

\bibitem{Watanabe2015} Y.X. Watanabe, Y.H. Kim, S.C. Jeong, Y. Hirayama, N. Imai, H. Ishiyama, H.S. Jung, H. Miyatake, S. Choi, J.S. Song, E. Clement, G. de France, A. Navin, M. Rejmund, C. Schmitt, G. Pollarolo, L. Corradi, E. Fioretto, D. Montanari, M. Niikura, D. Suzuki, H. Nishibata, and J. Takatsu, Phys. Rev. Lett. {\bf 115}, 172503 (2015).

\bibitem{Hirayama2017} Y. Hirayama, M. Mukai, Y.X. Watanabe, M. Ahmed, S.C. Jeong, H.S. Jung, Y. Kakiguchi, S. Kanaya, S. Kimura, J. Y. Moon, T. Nakatsukasa, M. Oyaizu, J.H. Park, P. Schury, A. Taniguchi, M. Wada, K. Washiyama, H. Watanabe, and H. Miyatake, Phys. Rev. C {\bf 96}, 014307 (2017).


\bibitem{Chen2018-1} J. Chen, J.L. Lou, Y.L. Ye, Z.H. Li, D.Y. Pang, C.X. Yuan, Y.C. Ge, Q.T. Li, H. Hua, D.X. Jiang, X.F. Yang, F.R. Xu, J.C. Pei, J. Li, W. Jiang, Y.L. Sun, H.L. Zang, Y. Zhang, N. Aoi, E. Ideguchi, H.J. Ong, J. Lee, J. Wu, H.N. Liu, C. Wen, Y. Ayyad, K. Hatanaka, T.D. Tran, T. Yamamoto, M. Tanaka, T. Suzuki, Phys. Lett. B {\bf 781}, 412 (2018).
\bibitem{Chen2018-2} J. Chen, J.L. Lou, Y.L. Ye, Z.H. Li, D.Y. Pang, C.X. Yuan, Y.C. Ge, Q.T. Li, H. Hua, D.X. Jiang, X.F. Yang, F.R. Xu, J.C. Pei, J. Li, W. Jiang, Y.L. Sun, H.L. Zang, Y. Zhang, G. Li, N. Aoi, E. Ideguchi, H.J. Ong, J. Lee, J. Wu, H.N. Liu, C. Wen, Y. Ayyad, K. Hatanaka, D.T. Tran, T. Yamamoto, M. Tanaka, and T. Suzuki, Phys Rev. C {\bf 98}, 014616 (2018).
\bibitem{Han2017} R. Han, X.Q. Li, W.G. Jiang, Z.H. Li, H. Hua, S.Q. Zhang, C.X. Yuan, D.X. Jiang, Y.L. Ye, J. Li, Z.H. Li, F.R. Xu, Q.B. Chen, J. Meng, J.S. Wang, C. Xu, Y.L. Sun, C.G. Wang, H.Y. Wu, C.Y. Niu, C.G. Li, C. He, W. Jiang, P.J. Li, H.L. Zang, J. Feng, S.D. Chen, Q. Liu, X.C. Chen, H.S. Xu, Z.G. Hu, Y.Y. Yang, P. Ma, J.B. Ma, S.L. Jin, Z. Bai, M.R. Huang, Y.J. Zhou, W.H. Ma, Y. Li, X.H. Zhou, Y.H. Zhang, G.Q. Xiao, W.L. Zhan, Phys. Lett. B {\bf 772}, 529 (2017).
\bibitem{Chen2019} Z.Q. Chen, Z.H. Li, H. Hua, H. Watanabe, C.X. Yuan, S.Q. Zhang, G. Lorusso, S. Nishimura, H. Baba, F. Browne, G. Benzoni, K. Y. Chae, F.C.L. Crespi, P. Doornenbal, N. Fukuda, G. Gey, R. Gernh\"{a}user, N. Inabe, T. Isobe, D.X. Jiang, A. Jungclaus, H.S. Jung, Y. Jin, D. Kameda, G.D. Kim, Y.K. Kim, I. Kojouharov, F.G. Kondev, T. Kubo, N. Kurz, Y.K. Kwon, X.Q. Li, J.L. Lou, G.J. Lane, C.G. Li, D.W. Luo, A. Montaner-Piz\'{a}, K. Moschner, C.Y. Niu, F. Naqvi, M. Niikura, H. Nishibata, A. Odahara, R. Orlandi, Z. Patel, Zs. Podoly\'{a}k, T. Sumikama, P.-A. S\"{o}derstr\"{o}m, H. Sakurai, H. Schaffner, G.S. Simpson, K. Steiger, H. Suzuki, J. Taprogge, H. Takeda, Zs. Vajta, H.K. Wang, J. Wu, A. Wendt, C.G. Wang, H.Y. Wu, X. Wang, C.G. Wu, C. Xu, Z.Y. Xu, A. Yagi, Y.L. Ye, and K. Yoshinaga, Phys. Rev. Lett. {\bf 122}, 212502 (2019).

\bibitem{Steer2008} S.J. Steer, Zs. Podoly\'{a}k, S. Pietri, M.G\'{o}rska, P.H. Regan, D. Rudolph, E. Werner-Malento, A.B. Garnsworthy, R. Hoischen, J. Gerl, H.J. Wollersheim, K.H. Maier, H. Grawe, F. Becker, P. Bednarczyk, L. C\'{a}ceres, P. Doornenbal, H. Geissel, J. Gr\c{e}bosz, A. Kelic, I. Kojouharov, N. Kurz, F. Montes, W. Prokopwicz, T. Saito, H. Schaffner, S. Tashenov, A. Heinz, M. Pf\"{u}tzner, T. Kurtukian-Nieto, G. Benzoni, A. Jungclaus, D.L. Balabanski, C. Brandau, A. Brown, A.M. Bruce, W.N. Catford, I.J. Cullen, Zs. Dombr\'{a}di, M.E. Estevez, W. Gelletly, G. Ilie, J. Jolie, G.A. Jones, M. Kmiecik, F.G. Kondev, R. Kr\"{u}cken, S. Lalkovski, Z. Liu, A. Maj, S. Myalski, S. Schwertel, T. Shizuma, P. M. Walker, E. Werner-Malento, and O. Wieland, Phys. Rev. C {\bf 78}, 061302(R) (2008).

\bibitem{nndc} http://www.nndc.bnl.gov/nudat3/



\bibitem{Wilson2015} E. Wilson, Zs. Podoly\'{a}k, H. Grawe, B.A. Brown, C.J. Chiara, S. Zhu, B. Fornal, R.V.F. Janssens, C.M. Shand, M. Bowry, M. Bunce, M.P. Carpenter, N. Cieplicka-Ory\'{n}czak, A.Y. Deo, G.D. Dracoulis, C.R. Hoffman, R.S. Kempley, F.G. Kondev, G.J. Lane, T. Lauritsen, G. Lotay, M.W. Reed, P.H. Regan, C. Rodr\'{i}guez Triguero, D. Seweryniak, B. Szpak, P.M. Walker, Phys. Lett. B {\bf 747}, 88 (2015).

\bibitem{Suzuki2012} T. Suzuki, T. Yoshida, T. Kajino, T. Otsuka, Phys. Rev. C {\bf 85}, 15802 (2012).
\bibitem{Nishimura2016} N. Nishimura, Zsolt Podoly\'{a}k, D.L. Fang, T. Suzuki, Phys. Lett. B {\bf 756}, 273 (2016).

\bibitem{AME2020} M. Wang, W.J. Huang, F.G. Kondev, G. Audi, S. Naimi, Chin. Phys. C, {\bf 45}(3), 030003 (2021).

\bibitem{Warburton1991-1} E.K. Warburton, Phys. Rev. C {\bf 44}, 233 (1991).
\bibitem{Kuo1971} T.T.S. Kuo and G.H. Herling, US Naval Research Laboraotory, Report No. 2258, unpublished (1971).
\bibitem{Herling1972} G.H. Herling and T.T.S. Kuo, Nucl. Phys. {\bf A 181}, 113 (1972).
\bibitem{Rydstrom1990} L. Rydstr\"{o}m, J. Blomqvist, R. J. Liotta, and C. Pomar, Nucl. Phys. {\bf A 512}, 217 (1990).

\bibitem{Warburton1991-2} E.K. Warburton and B.A. Brown, Phys. Rev. C {\bf 43}, 602 (1991).

\bibitem{sun2017PLB} M.D. Sun, Z. Liu, T.H. Huang, W.Q. Zhang, J.G. Wang, X.Y. Liu, B. Ding, Z.G. Gan, L. Ma, H.B. Yang, Z.Y. Zhang, L. Yu, J. Jiang, K.L. Wang, Y.S. Wang, M.L. Liu, Z.H. Li, J. Li, X. Wang, H.Y. Lu, C.J. Lin, L.J. Sun, N.R. Ma, C.X. Yuan, W. Zuo, H.S. Xu, X.H. Zhou, G.Q. Xiao, C. Qi, and F.S. Zhang, Phys. Lett. B {\bf 771}, 303 (2017).

\bibitem{zhang2020PLB} M.M. Zhang, H.B. Yang, Z.G. Gan, Z.Y. Zhang, M.H. Huang, L. Ma, C.L. Yang, C.X. Yuan, Y.S. Wang, Y.L.Tian, H.B. Zhou, S. Huang, X.T. He, S.Y. Wang, W.Z. Xu, H.W. Li, X.X. Xu, J.G. Wang, H.R. Yang, L.M. Duan, W.Q. Yang, S.G. Zhou, Z.Z. Ren, X.H. Zhou, H.S. Xu, A.A. Voinovk, Yu.S. Tsyganovk, A.N. Polyakovk, and M.V. Shumeiko, Phys. Lett. B {\bf 800}, 135102 (2020).

\bibitem{otsuka2010} T. Otsuka, T. Suzuki, M. Honma, Y. Utsuno, N. Tsunoda, K. Tsukiyama, and M. Hjorth-Jensen, Phys. Rev. Lett. {\bf 104}, 012501 (2010).
\bibitem{m3y1977} G. Bertsch, J. Borysowicz, H. McManus, and W. G. Love, Nucl. Phys. {\bf A284}, 399 (1977).

\bibitem{yuan2012} C. Yuan, T. Suzuki, T. Otsuka, F.R. Xu, and N. Tsunoda, Phys Rev. C {\bf 85}, 064324 (2012).
\bibitem{utsuno2012} Y. Utsuno, T. Otsuka, B. A. Brown, M. Honma, T. Mizusaki, and N. Shimizu, Phys Rev. C {\bf 86}, 051301(R) (2012).
\bibitem{Togashi2015} T. Togashi, N. Shimizu, Y. Utsuno, T. Otsuka, and M. Honma, Phys Rev. C {\bf 91}, 024320 (2015).

\bibitem{yuan2016} C. Yuan, Zhong Liu, Furong Xu, P.M. Walker, Zs. Podoly\'{a}k, C. Xu, Z.Z. Ren, B. Ding, M.L. Liu, X.Y. Liu, H.S. Xu, Y.H. Zhang, X.H. Zhou, W. Zuo, Phys. Lett. B {\bf 762}, 237 (2016).

\bibitem{zhang2021-U214} Z.Y. Zhang, H.B. Yang, M.H. Huang, Z.G. Gan, C.X. Yuan, C. Qi, A.N. Andreyev, M.L. Liu, L. Ma, M.M. Zhang, Y.L. Tian, Y.S. Wang, J.G. Wang, C.L. Yang, G.S. Li, Y.H. Qiang, W.Q. Yang, R.F. Chen, H.B. Zhang, Z.W. Lu, X.X. Xu, L.M. Duan, H.R. Yang, W.X. Huang, Z. Liu, X.H. Zhou, Y.H. Zhang, H.S. Xu, N. Wang, H.B. Zhou, X.J. Wen, S. Huang, W. Hua, L. Zhu, X. Wang, Y.C. Mao, X.T. He, S.Y. Wang, W.Z. Xu, H.W. Li, Z.Z. Ren and S.G. Zhou, Phys. Rev. Lett. {\bf 126}, 152502 (2021).

\bibitem{yang2022-Th207} H.B. Yang, Z.G. Gan, Z.Y. Zhang, M.H. Huang, L. Ma, M.M. Zhang, C.X. Yuan, Y.F. Niu, C.L. Yang, Y.L. Tian, L. Guo, Y.S. Wang, J.G. Wang, H.B. Zhou, X.J. Wen, H.R. Yang, X.H. Zhou, Y.H. Zhang, W.X. Huang, Z. Liu, S.G. Zhou, Z.Z. Ren, H.S. Xu, V.K. Utyonkov, A.A. Voinov, Y.S. Tsyganov, A.N. Polyakov and D.I. Solovyev, Phys Rev. C {\bf 105}, L051302 (2022).


\bibitem{yuan2017} C.X. Yuan, Chin. Phys. C, {\bf41}(10), 104102, (2017).


\bibitem{myers1966} W.D. Myers and W.J. Swiatecki, Nucl. Phys. {\bf 81}, 1 (1966).
\bibitem{moller1995} P. M\"{o}ller and J.R. Nix, At. Data Nucl. Data Tables  {\bf 59}, 185 (1995).
\bibitem{pomorski2003} K. Pomorski and J. Dudek, Phys. Rev. C {\bf 67}, 044316 (2003).



\bibitem{Goriely2009} S. Goriely, N. Chamel, and J.M. Pearson, Phys. Rev. Lett. {\bf 102}, 152503 (2009).
\bibitem{Goriely2013} S. Goriely, N. Chamel, and J.M. Pearson, Phys. Rev. C {\bf 88}, 061302(R) (2013).

\bibitem{liu2011} M. Liu, N.Wang, Y. Deng, and X. Wu, Phys. Rev. C {\bf 84}, 014333 (2011).
\bibitem{Wang2014} N. Wang, M. Liu, X.Z. Wu, and J. Meng, Phys. Lett. B {\bf 734}, 215 (2014).
\bibitem{Suzuki2016} T. Suzuki, T. Otsuka, C.X. Yuan, and A. Navin, Phys. Lett. B {\bf 753}, 199 (2016).

\bibitem{garvey1966} G.T. Garvey and I. Kelson, Phys. Rev. Lett. {\bf 16}, 197 (1966).

\bibitem{cheng2014} Y.Y. Cheng, Y.M. Zhao, and A. Arima, Phys. Rev. C {\bf 90}, 064304 (2014).

\bibitem{fu2018} G.J. Fu, Y.Y. Cheng, Y.H. Zhang, P. Zhang, P. Shuai, Y.M. Zhao, and M. Wang, Phys. Rev. C {\bf 97}, 024339 (2018).

\bibitem{Amro2017} B.M.S. Amro, C.J. Lister, E.A. McCutchan, W. Loveland, P. Chowdhury, S. Zhu, A.D. Ayangeakaa, J.S. Barrett, M.P. Carpenter, C.J. Chiara, J.P. Greene, J.L. Harker, R.V.F. Janssens, T. Lauritsen, A.A. Sonzogni, W.B. Walters, and R. Yanez, Phys. Rev. C {\bf 95}, 014330 (2017).

\bibitem{Gottardo2019} A. Gottardo, J.J. Valiente-Dob\'{o}n, G. Benzoni, A.I. Morales, A. Gadea, S. Lunardi, P. Boutachkov, A.M. Bruce, M. G\'{o}rska, J. Grebosz, S. Pietri, Zs. Podoly\'{a}k, M. Pf\"{u}tzner, P.H. Regan, D. Rudolph, H. Weick, J. Alc\'{a}ntara N\'{u}\~{n}ez, A. Algora, N. Al-Dahan, G. de Angelis, Y. Ayyad, N. Alkhomashi, P.R.P. Allegro, D. Bazzacco, J. Benlliure, M. Bowry, A. Bracco, M. Bunce, F. Camera, E. Casarejos, M.L. Cortes, F.C.L. Crespi, A. Corsi, A.M. Denis Bacelar, A.Y. Deo, C. Domingo-Pardo, M. Doncel, Zs. Dombradi, T. Engert, K. Eppinger, G.F. Farrelly, F. Farinon, H. Geissel, J. Gerl, N. Goel, E. Gregor, T. Habermann, R. Hoischen, R. Janik, S. Klupp, I. Kojouharov, N. Kurz, S.M. Lenzi, S. Leoni, S. Mandal, R. Menegazzo, D. Mengoni, B. Million, D.R. Napoli, F. Naqvi, C. Nociforo, A. Prochazka, W. Prokopowicz, F. Recchia, R.V. Ribas, M.W. Reed, E. Sahin, H. Schaffner, A. Sharma, B. Sitar, D. Siwal, K. Steiger, P. Strmen, T.P.D. Swan, I. Szarka, C.A. Ur, P.M. Walker, O. Wieland, H-J. Wollersheim, F. Nowacki, and E. Maglione, Phys. Rev. C {\bf 99}, 054326 (2019).

\bibitem{Broda2018}R. Broda, \L.W. Iskra, R.V.F. Janssens, B.A. Brown, B. Fornal, J. Wrzesi\'{n}ski, N. Cieplicka-Ory\'{n}czak, M.P. Carpenter, C.J. Chiara, C.R. Hoffman, F.G. Kondev, G.J. Lane, T. Lauritsen, Zs. Podoly\'{a}k, D. Seweryniak, W. B. Walters, and S. Zhu, Phys. Rev. C {\bf 98}, 024324 (2018).

\bibitem{Podolyak2009} Zs. Podoly\'{a}k, G.F. Farrelly, P.H. Regan, A.B. Garnsworthy, S.J. Steer, M. G\'{o}rska, J. Benlliure, E. Casarejos, S. Pietri, J. Gerl, H.J. Wollersheim, R. Kumar, F. Molina, A. Algora, N. Alkhomashi, G. Benzoni, A. Blazhev, P. Boutachkov, A.M. Bruce, L. Caceres, I.J. Cullen, A.M. Denis Bacelar, P. Doornenbal, M.E. Estevez, Y. Fujita, W. Gelletly, H. Geissel, H. Grawe, J. Gr\c{e}bosz, R. Hoischen, I. Kojouharov, S. Lalkovski, Z. Liu, K.H. Maier, C. Mihai, D. M\"{u}cher, B. Rubio, H. Schaffner, A. Tamii, S. Tashenov, J.J. Valiente-Dob\'{o}n, P.M. Walker, P.J. Woods, Phys. Lett. B {\bf 672}, 116 (2009).

\bibitem{Podolyak2009-2} Zs. Podoly\'{a}k, S.J. Steer, S. Pietri, M. G\'{o}rska, P.H. Regan, D. Rudolph, A.B. Garnsworthy, R. Hoischen, J. Gerl, H.J. Wollersheim, H. Grawe, K.H. Maier, F. Becker, P. Bednarczyk, L. C\'{a}ceres, P. Doornenbal, H. Geissel, J. Grebosz, A. Kelic, I. Kojouharov, N. Kurz, F. Montes, W. Prokopowicz, T. Saito, H. Schaffner, S. Tashenov, A. Heinz, T. Kurtukian-Nieto, G. Benzoni, M. Pf\"{u}tzner, A. Jungclaus, D.L. Balabanski, C. Brandau, B.A. Brown, A.M. Bruce, W.N. Catford, I.J. Cullen, Zs. Dombr\'{a}di, M.E. Estevez, W. Gelletly, G. Ilie, J. Jolie, G.A. Jones, M. Kmiecik, F.G. Kondev, R. Kr\"{u}cken, S. Lalkovski, Z. Liu, A. Maj, S. Myalski, S. Schwertel, T. Shizuma, P.M. Walker, E. Werner-Malento, O. Wieland, Eur. Phys. J. A {\bf 42}, 489 (2009).

\bibitem{John2017} P.R. John, J.J. Valiente-Dob\'{o}n, D. Mengoni, V. Modamio, S. Lunardi, D. Bazzacco, A. Gadea, C. Wheldon, T.R. Rodr\'{i}guez, T. Alexander, G. de Angelis, N. Ashwood, M. Barr, G. Benzoni, B. Birkenbach, P.G. Bizzeti, A.M. Bizzeti-Sona, S. Bottoni, M. Bowry, A. Bracco, F. Browne, M. Bunce, F. Camera, L. Corradi, F.C.L. Crespi, B. Melon, E. Farnea, E. Fioretto, A. Gottardo, L. Grente, H. Hess, Tz. Kokalova, W. Korten, A. Ku\c{s}o\u{g}lu, S. Lenzi, S. Leoni, J. Ljungvall, R. Menegazzo, C. Michelagnoli, T. Mijatovi\'{c}, G. Montagnoli, D. Montanari, D. R. Napoli, Zs. Podoly\'{a}k, G. Pollarolo, F. Recchia, P. Reiter, O.J. Roberts, E. \c{S}ahin, M.-D. Salsac, F. Scarlassara, M. Sferrazza, P.-A. S\"{o}derstr\"{o}m, A.M. Stefanini, S. Szilner, C.A. Ur, A. Vogt, and J. Walshe, Phys. Rev. C {\bf 95}, 064321 (2017).

\bibitem{Jiang2011-1} H. Jiang, J.J. Shen, Y.M. Zhao, and A Arima, J. Phys. G: Nucl. Part. Phys. {\bf 38}, 045103 (2011).
\bibitem{Jiang2011-2} H. Jiang and Y.M. Zhao, Sci. China Phys. Mech. Astron. {\bf 54}, 1461 (2011).

\bibitem{otsuka2001} T. Otsuka, R. Fujimoto, Y. Utsuno, B. A. Brown, M. Honma, and T. Mizusaki, Phys. Rev. Lett. {\bf 87}, 082502 (2001).
\bibitem{otsuka2005} T. Otsuka, T. Suzuki, R. Fujimoto, H. Grawe, and Y. Akaishi, Phys. Rev. Lett. {\bf 95}, 232502 (2005).
\bibitem{Xu2019} X. Xu, J.H. Liu, C.X. Yuan, Y.M. Xing, M. Wang, Y.H. Zhang, X.H. Zhou, Yu. A. Litvinov, K. Blaum, R.J. Chen, X.C. Chen, C.Y. Fu, B.S. Gao, J.J. He, S. Kubono, Y.H. Lam, H.F. Li, M.L. Liu, X.W. Ma, P. Shuai, M. Si, M.Z. Sun, X.L. Tu, Q. Wang, H.S. Xu, X.L. Yan, J.C. Yang, Y.J. Yuan, Q. Zeng, P. Zhang, X. Zhou, W.L. Zhan, S. Litvinov, G. Audi, S. Naimi, T. Uesaka, Y. Yamaguchi, T. Yamaguchi, A. Ozawa, B.H. Sun, K. Kaneko, Y. Sun, and F.R. Xu, Phys. Rev. C {\bf 100}, 051303(R) (2019).

\bibitem{Rejmund1999} M. Rejmund, M. Schramm and K.H. Maier, Phys. Rev. C {\bf 59}, 2520 (1999).

\bibitem{lane1960} A.M. Lane and E.D. Pendlebury, Nucl. Phys. {\bf 15}, 39 (1960).
\bibitem{carter1960} J.C. Carter, W.T. Pinkston, and W.W. True, Phys. Rev. {\bf 120}, 504 (1960).

\bibitem{Stone2016} N.J. Stone, At. Data Nucl. Data Tables {\bf 111}, 1 (2016).
\bibitem{BrIcc} T. Kib\'{e}di, T.W. Burrows, M.B. Trzhaskovskaya, P.M. Davidson, C.W. Nestor Jr., Nucl. Instr. and Meth. A {\bf 589}, 202 (2008); http://bricc.anu.edu.au

\bibitem{Richter2008PRC} W.A. Richter, S. Mkhize, and B.A. Brown, Phys. Rev. C {\bf78}, 064302 (2008)

\bibitem{Honma2009} M. Honma, T. Otsuka, T. Mizusaki, and M. Hjorth-Jensen, Phys. Rev. C {\bf80}, 064323 (2009)

\bibitem{arima1986} A. Arima, K. Shimizu, W. Bentz, and H. Hyuga, Adv. Nucl.
Phys. {\bf 18}, 1 (1986); A. Arima and H. Hyuga, in \emph{Mesons in
Nuclei}, edited by D. H. Wilkinson and M. Rho (North-Holland,
Amsterdam, 1979), Vol. II, p. 683.
\bibitem{towner1987} I. S. Towner, Phys. Rep. {\bf 155}, 263 (1987); I. S. Towner and F. C. Khanna, Nucl. Phys. {\bf A399}, 334 (1983).

\bibitem{Stone2005} N.J. Stone, At. Data Nucl. Data Tables {\bf 90}, 75 (2005).

\bibitem{Morales2014} A.I. Morales, G. Benzoni, A. Gottardo, J.J. Valiente-Dob\'{o}n, N. Blasi, A. Bracco, F. Camera, F.C.L. Crespi, A. Corsi, S. Leoni, B. Million, R. Nicolini, O. Wieland, A. Gadea, S. Lunardi, M. G\'{o}rska, P.H. Regan, Zs. Podoly\'{a}k, M. Pf\"{u}tzner, S. Pietri, P. Boutachkov, H. Weick, J. Grebosz, A.M. Bruce, J. Alc\'{a}ntara N\'{u}\~{n}ez, A. Algora, N. Al-Dahan, Y. Ayyad, N. Alkhomashi, P.R.P. Allegro, D. Bazzacco, J. Benlliure, M. Bowry, M. Bunce, E. Casarejos, M.L. Cortes, A.M. Denis Bacelar, A.Y. Deo, G. de Angelis, C. Domingo-Pardo, M. Doncel, Zs. Dombradi, T. Engert, K. Eppinger, G.F. Farrelly, F. Farinon, E. Farnea, H. Geissel, J. Gerl, N. Goel, E. Gregor, T. Habermann, R. Hoischen, R. Janik, S. Klupp, I. Kojouharov, N. Kurz, S. Mandal, R. Menegazzo, D. Mengoni, D.R. Napoli, F. Naqvi, C. Nociforo, A. Prochazka, W. Prokopowicz, F. Recchia, R.V. Ribas, M.W. Reed, D. Rudolph, E. Sahin, H. Schaffner, A. Sharma, B. Sitar, D. Siwal, K. Steiger, P. Strmen, T.P.D. Swan, I. Szarka, C.A. Ur, P.M. Walker, and H.-J. Wollersheim, Phys. Rev. C {\bf 89}, 014324 (2014).
\bibitem{Caballero2016} R. Caballero-Folch, C. Domingo-Pardo, J. Agramunt, A. Algora, F. Ameil, A. Arcones, Y. Ayyad, J. Benlliure, I.N. Borzov, M. Bowry, F. Calvi\~{n}o, D. Cano-Ott, G. Cort\'{e}s, T. Davinson, I. Dillmann, A. Estrade, A. Evdokimov, T. Faestermann, F. Farinon, D. Galaviz, A.R. Garc\'{i}a, H. Geissel, W. Gelletly, R. Gernh\"{a}user, M. B. G\'{o}mez-Hornillos, C. Guerrero, M. Heil, C. Hinke, R. Kn\"{o}bel, I. Kojouharov, J. Kurcewicz, N. Kurz, Yu. A. Litvinov, L. Maier, J. Marganiec, T. Marketin, M. Marta, T. Mart\'{i}nez, G. Mart\'{i}nez-Pinedo, F. Montes, I. Mukha, D.R. Napoli, C. Nociforo, C. Paradela, S. Pietri, Zs. Podoly\'{a}k, A. Prochazka, S. Rice, A. Riego, B. Rubio, H. Schaffner, Ch. Scheidenberger, K. Smith, E. Sokol, K. Steiger, B. Sun, J.L. Ta\'{i}n, M. Takechi, D. Testov, H. Weick, E. Wilson, J.S. Winfield, R. Wood, P. Woods, and A. Yeremin, Phys. Rev. Lett. {\bf 117}, 012501 (2016)
\bibitem{Caballero2017} R. Caballero-Folch, C. Domingo-Pardo, J. Agramunt, A. Algora, F. Ameil, Y. Ayyad, J. Benlliure, M. Bowry, F. Calvi\~{n}o, D. Cano-Ott, G. Cort\`{e}s, T. Davinson, I. Dillmann, A. Estrade, A. Evdokimov, T. Faestermann, F. Farinon, D. Galaviz, A.R. Garc\'{i}a, H. Geissel, W. Gelletly, R. Gernh\"{a}user, M.B.G\'{o}mez-Hornillos, C. Guerrero, M. Heil, C. Hinke, R. Kn\"{o}bel, I. Kojouharov, J. Kurcewicz, N. Kurz, Yu.A. Litvinov, L. Maier, J. Marganiec, M. Marta, T. Mart\'{i}nez, F. Montes, I. Mukha, D.R. Napoli, C. Nociforo, C. Paradela, S. Pietri, Zs. Podoly\'{a}k, A. Prochazka, S. Rice, A. Riego, B. Rubio, H. Schaffner, Ch. Scheidenberger, K. Smith, E. Sokol, K. Steiger, B. Sun, J.L. Ta\'{i}n, M. Takechi, D. Testov, H. Weick, E. Wilson, J.S. Winfield, R. Wood, P.J. Woods, and A. Yeremin, Phys. Rev. C {\bf 95}, 064322 (2017).

\end{thebibliography}
\end{document}